\documentclass[letterpaper,twocolumn]{jpsj3}
\usepackage{txfonts}
\usepackage{amsmath}
\usepackage{amssymb}
\usepackage{bm}
\newcommand{\rev}[1]{#1}
\newenvironment{revblock}{}{}

\title{Magnetic-Field Effects on the Electronic States of the K$_4$ Lattice}

\author{Katsuhiro Takahashi$^1$, Shouya Yoshida$^1$, and Katsunori Wakabayashi$^{2,1,3,*}$}
\inst{$^1$Department of Nanotechnology for Sustainable Energy, School of Science and Technology, \\
Kwansei Gakuin University, Gakuen-Uegahara 1, Sanda 669-1330, Japan \\
$^2$Research Center for Materials Nanoarchitectonics (MANA),\\
National Institute for Materials Science (NIMS), Namiki 1-1, Tsukuba 305-0044, Japan\\
$^3$Center for Spintronics Research Network (CSRN), Osaka University, Toyonaka 560-8531, Japan
}

\abst{
We investigate the electronic states of the K$_4$ lattice in a uniform magnetic field within a nearest-neighbor tight-binding model.
The K$_4$ lattice is a \rev{noncentrosymmetric three-dimensional network with triple-Dirac cones at the $\Gamma$ and H points}.
By fixing the Bloch momentum component parallel to the field, the Peierls-substituted tight-binding problem reduces to an effective two-dimensional problem and yields a Hofstadter spectrum \rev{whose self-similar subpatterns reflect the geometry of the lattice projected onto the plane perpendicular to the field}.
A square-like substructure emerges on the flux scale $\phi = 1/8$ for $\bm{B}\parallel[100]$, and honeycomb- and triangular-like substructures emerge on the finer scale $\phi = 1/32$ for $\bm{B}\parallel[111]$\rev{.
Here $\phi$ is the flux through the elementary plaquette of the projected lattice, and these scales correspond to one flux quantum per projected unit cell.
Owing to the screw axes of the chiral K$_4$ network, the field-parallel momentum acts as an additional flux, which, for commensurate values of this momentum, shifts the spectrum along the flux axis and brings the second triple-Dirac cone at the H point into the magnetic spectrum}.
In the weak-field limit, we derive the Landau-level structure from a low-energy effective Hamiltonian around the $\Gamma$ point.
The three-component nature of the triple-Dirac cone gives rise to a cubic secular equation, whose solutions describe how a pseudospin-flat branch coexists with two relativistic Landau ladders \rev{and a single chiral $n=0$ level dispersing linearly along the field, in agreement with the lattice calculation}.
}

\begin{document}
\maketitle

\section{Introduction}

The K$_4$ lattice is a three-dimensional periodic network derived from the complete graph $K_4$ in graph theory.\cite{Sunada2008,Sunada2012JJM,Sunada2013Book}
Its primitive unit cell contains four sublattices, and each site is connected to three equivalent neighbors at the same distance.
The same network is also referred to as the srs lattice or the Laves graph.\cite{OKeeffe2008,Hyde2008}
Because of this highly symmetric connectivity, the K$_4$ lattice has emerged as a clean model system in which the geometry of the network directly controls the electronic structure.

In terms of local bonding, the K$_4$ lattice can be regarded as a three-dimensional generalization of graphene.
Each site forms three coplanar bonds, as in $sp^2$-hybridized carbon, but the bond planes at neighboring sites are tilted relative to each other by $\cos^{-1}(1/3) \simeq 70.5^\circ$.
This twist produces a fully three-dimensional network that cannot be reduced to any stack of two-dimensional layers.
The associated symmetry forces a triple band crossing at both the $\Gamma$ and H points of the Brillouin zone, giving rise to triple-Dirac cones consisting of two linearly dispersing bands and one weakly dispersing branch.\cite{Tsuchiizu2016PRB,Yoshida2026PRB}
Such higher-pseudospin crossings, classified more generally as unconventional quasiparticles in noncentrosymmetric crystals,\cite{Bradlyn2016Science,Armitage2018RMP} have recently been proposed in WC-type compounds\cite{Weng2016PRB} and experimentally observed at the triply degenerate nodal points of the topological semimetal MoP.\cite{Lv2017Nature}
These developments place the K$_4$ lattice alongside Weyl semimetals\cite{Burkov2011PRL,Lv2015PRX} and Dirac semimetals as a target for studying topological response and magnetic spectroscopy in three dimensions.

Possible material realizations of K$_4$-type networks have been actively explored.
While the original carbon-based K$_4$ structure was shown to be dynamically unstable,\cite{Itoh2009PRL,Hagita2016JCCS} K$_4$-type connectivity has been predicted for boron, phosphorus, hydrogenated carbon, carbon--nitrogen systems, and Na--C compounds.\cite{Dai2010PCCP,Liu2016SR,Lian2013JCP,Wen2011CPL,Wen2012JPCS}
Most notably, the K$_4$ network has been realized experimentally in a molecular crystal formed by a triangular $\pi$-radical anion.\cite{Mizuno2015JACS}
These developments establish the K$_4$ structure as a viable electronic motif, independent of the metastability of pristine K$_4$ carbon, and motivate quantitative studies of its response to external fields.

A magnetic field provides a particularly informative probe.
In two-dimensional periodic systems with rational flux per unit cell, Bloch electrons reorganize into the Hofstadter spectrum, whose self-similar structure encodes the lattice symmetry and topology.\cite{Peierls1933ZPhys,Harper1955,Brown1964PR,Zak1964PR,Hofstadter1976PRB,Thouless1982PRL,Kohmoto1985AnnPhys,Hasegawa1989PRL,Janecek2013PRB}
Recent experiments on moir\'e superlattices have brought the butterfly into the laboratory.\cite{Dean2013Nature,Ponomarenko2013Nature}
For a three-dimensional crystal, fixing the Bloch momentum component along the field reduces the magnetic Schr\"{o}dinger equation to an effective two-dimensional problem.
The Hofstadter spectrum then becomes a function of \rev{the field-parallel momentum} as well as of flux, and the projected lattice geometry directly imprints itself on the spectrum.\cite{KoshinoAoki2001PRL}
The K$_4$ lattice, with its unconventional symmetry and triple-Dirac low-energy structure, is a natural three-dimensional platform for this analysis.

The triple-Dirac point at $\Gamma$ likewise motivates a complementary weak-field analysis.
Starting from a $3\times 3$ low-energy effective Hamiltonian\cite{Tsuchiizu2016PRB} and applying minimal coupling, one can derive the corresponding Landau-level spectrum analytically.
For the K$_4$ triple-Dirac Hamiltonian, this construction leads to a cubic secular equation that quantitatively describes how Landau levels emerge from the threefold crossing\cite{Landau1930ZPhys,McClure1956PR,Dresselhaus1974PRB,McCann2006PRL,Goerbig2011RMP,Raoux2014PRL,Malcolm2014PRB,Illes2016PRB} and that accounts for the low-flux behavior of the numerical Hofstadter spectrum.

In this paper, we provide a unified study of magnetic-field effects in the K$_4$ lattice.
After establishing our conventions for the tight-binding model in Sect.~2, we present in Sect.~3 the Hofstadter spectrum for $\bm{B}\parallel[100]$ and $\bm{B}\parallel[111]$\rev{,} identify the characteristic flux scales associated with the projected plaquettes\rev{, and show that the field-parallel momentum enters the spectrum as an additional flux (spectra at finite field-parallel momentum are collected in the Appendix)}.
In Sect.~4 we derive the Landau-level spectrum from the low-energy effective Hamiltonian by two equivalent methods and verify\rev{, by a direct comparison with the lattice spectrum at finite field-parallel momentum,} that it consistently describes the weak-field limit of the full numerical calculation.
We close with a summary in Sect.~5.

\begin{figure*}[t]
\centering
\includegraphics[width=\textwidth,clip]{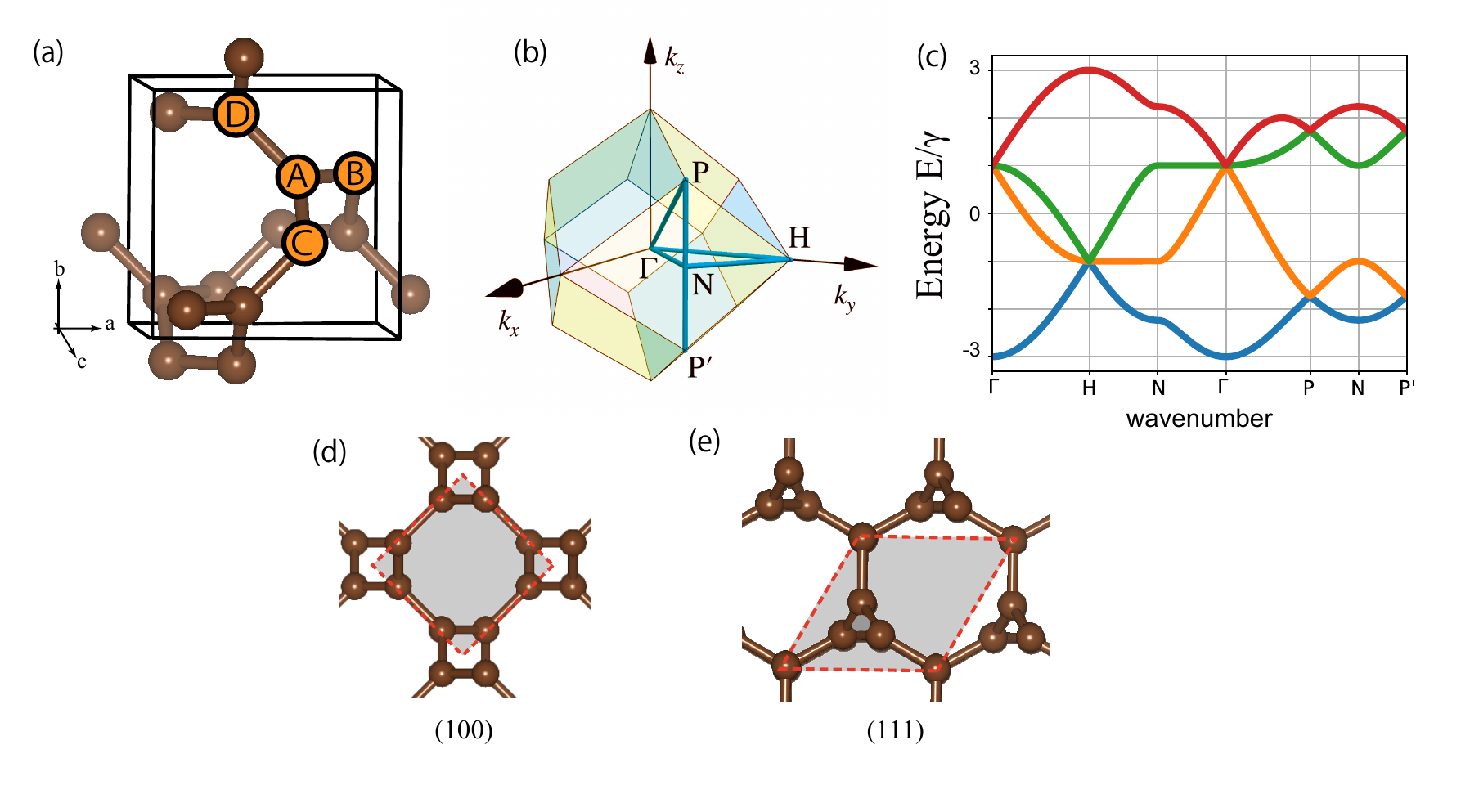}
\caption{
(Color online)
(a) Conventional cubic unit cell of the K$_4$ lattice with the four sublattices A, B, C, and D.
In a carbon realization, each site forms three $\sigma$ bonds via $sp^2$ hybridization, with the bond planes tilted by $\cos^{-1}(1/3)\simeq 70.5^\circ$ at neighboring sites.
(b) First Brillouin zone of the K$_4$ lattice with high-symmetry points indicated.
(c) Nearest-neighbor tight-binding band structure along the high-symmetry path $\Gamma$--H--N--$\Gamma$--P--N--P$'$ indicated in (b).
Triple band crossings appear at the $\Gamma$ ($E=+\gamma$) and H ($E=-\gamma$) points, forming the K$_4$ triple-Dirac cones, while doubly degenerate pairs occur at the P points.
The vertical axis is in units of the nearest-neighbor hopping amplitude $\gamma$, distinct from the energy--length factor $\gamma\,a=2\hbar v_F$ ($a$ is the lattice constant) that enters the effective Hamiltonian in Sect.~4.
(d) Lattice viewed along [100].
The dashed square marks the projected unit cell and reveals alternating eight- and four-membered rings.
(e) Lattice viewed along [111], showing a honeycomb sublattice decorated with a triangular sublattice.
}
\label{fig:k4_structure}
\end{figure*}

\section{Structure and Tight-Binding Model}

In this section we establish our conventions for the K$_4$ crystal structure and introduce the tight-binding Hamiltonian used throughout the paper.
Figure~\ref{fig:k4_structure}(a) shows the conventional cubic unit cell of the K$_4$ crystal.
The primitive cell contains four inequivalent sublattices, A, B, C, and D, each connected to three neighbors at the same distance, forming a fully three-dimensional network.

The K$_4$ crystal belongs to the noncentrosymmetric body-centered cubic space group $I4_132$, whose point group is the chiral octahedral group $O$.\cite{Yoshida2026PRB}
We adopt the rhombohedral primitive unit cell, with the conventional cubic lattice constant $a$.
The three primitive translation vectors read
\begin{align}
\bm{a}_1 &= \tfrac{a}{2}(-1,1,1), &
\bm{a}_2 &= \tfrac{a}{2}(1,-1,1), &
\bm{a}_3 &= \tfrac{a}{2}(1,1,-1).
\label{eq:primitive_vec}
\end{align}
The six nearest-neighbor bond vectors generated by the K$_4$ connectivity are
\begin{align}
\bm{t}_1 &= \tfrac{a}{4}(0,1,-1), &
\bm{t}_2 &= \tfrac{a}{4}(-1,0,1), &
\bm{t}_3 &= \tfrac{a}{4}(1,-1,0), \nonumber \\
\bm{t}_4 &= \tfrac{a}{4}(1,1,0), &
\bm{t}_5 &= \tfrac{a}{4}(-1,0,-1), &
\bm{t}_6 &= \tfrac{a}{4}(0,1,1),
\label{eq:bondvec}
\end{align}
all of which share the same length $|\bm{t}_n| = a/(2\sqrt{2})$.
Following the convention of Ref.~\citen{Yoshida2026PRB}, we place the four sublattice sites of the primitive cell at
\begin{align}
\bm{R}_A &= a(0,0,0), &
\bm{R}_B &= a\left(\tfrac{1}{2},\tfrac{1}{4},-\tfrac{1}{4}\right), \nonumber\\
\bm{R}_C &= a\left(-\tfrac{1}{4},\tfrac{1}{2},\tfrac{1}{4}\right), &
\bm{R}_D &= a\left(\tfrac{1}{4},-\tfrac{1}{4},\tfrac{1}{2}\right) ,
\label{eq:sublat}
\end{align}
which correspond to the Wyckoff position 8\textit{a} of the space group $I4_132$.
The K$_4$ nearest neighbors of any given site lie in adjacent primitive cells.
The six nearest-neighbor connections per primitive cell are generated by the bond vectors $\bm{t}_1$--$\bm{t}_6$ combined with the appropriate primitive translations.

Although the local bonding around each site echoes the $sp^2$ pattern of graphene---three bonds in a common plane---the bond planes are no longer coplanar.
At a neighboring site the plane is rotated by
\begin{equation}
\theta = \cos^{-1}(1/3) \simeq 70.5^\circ ,
\label{eq:twist}
\end{equation}
relative to the original plane.
Setting $\theta = 0$ would collapse the network to the two-dimensional honeycomb lattice.
The finite $\theta$ characteristic of K$_4$ produces a chiral three-dimensional structure containing ten-membered rings.\cite{Sunada2008,Sunada2013Book}
Figures~\ref{fig:k4_structure}(d) and (e) display the lattice viewed along [100] and [111], respectively.
In the (100) view, alternating four- and eight-membered rings tile the projected plane.
In the (111) view, the network resembles a honeycomb lattice decorated with a triangular sublattice.
These geometric features will reappear in Sect.~3 as the characteristic flux scales of the Hofstadter spectrum.

The corresponding reciprocal lattice vectors are
\begin{align}
\bm{b}_1 &= \frac{2\pi}{a}(0,1,1), &
\bm{b}_2 &= \frac{2\pi}{a}(1,0,1), &
\bm{b}_3 &= \frac{2\pi}{a}(1,1,0),
\end{align}
and the first Brillouin zone defined by these vectors is shown in Fig.~\ref{fig:k4_structure}(b).

We describe the electronic structure with a nearest-neighbor tight-binding model in which each site carries a single orbital and electrons hop to adjacent sites with amplitude $\gamma$.
In the sublattice basis $(A,B,C,D)$, the Bloch Hamiltonian reads
\begin{equation}
H(\bm{k})=
-\gamma\begin{pmatrix}
0 & e^{i\bm{k}\cdot\bm{t}_1} & e^{i\bm{k}\cdot\bm{t}_2} & e^{i\bm{k}\cdot\bm{t}_3} \\
e^{-i\bm{k}\cdot\bm{t}_1} & 0 & e^{i\bm{k}\cdot\bm{t}_4} & e^{i\bm{k}\cdot\bm{t}_5} \\
e^{-i\bm{k}\cdot\bm{t}_2} & e^{-i\bm{k}\cdot\bm{t}_4} & 0 & e^{i\bm{k}\cdot\bm{t}_6} \\
e^{-i\bm{k}\cdot\bm{t}_3} & e^{-i\bm{k}\cdot\bm{t}_5} & e^{-i\bm{k}\cdot\bm{t}_6} & 0
\end{pmatrix}.
\label{eq:HBloch}
\end{equation}
Diagonalizing Eq.~(\ref{eq:HBloch}) along the high-symmetry path indicated in Fig.~\ref{fig:k4_structure}(b) yields the energy eigenvalues $E$ that constitute the band structure of Fig.~\ref{fig:k4_structure}(c).
At the $\Gamma$ point, two linearly dispersing bands and a third weakly dispersing branch meet at a single energy, forming a triple-Dirac cone.
A threefold crossing also occurs at H, with the spectrum inverted relative to $\Gamma$.\cite{Tsuchiizu2016PRB,Yoshida2026PRB}
These crossings provide the starting point for the low-energy effective description used in the weak-field analysis of Sect.~4.

\begin{revblock}
Two further properties of the band structure will be used below.
First, the K$_4$ lattice is bipartite.
The sites $\bm{R}_\alpha+\sum_i n_i\bm{a}_i$ with $\sum_i n_i$ even form one sublattice and those with $\sum_i n_i$ odd the other, and every nearest-neighbor bond connects the two sublattices.
Because a single primitive translation exchanges the two sublattices, the associated chiral (sublattice) symmetry involves a momentum shift by the H point, $\bm{K}_{\mathrm H}=(2\pi/a)(1,0,0)$ or any of its equivalents,
\begin{equation}
E(\bm{k}+\bm{K}_{\mathrm H})=-E(\bm{k}).
\label{eq:chiral}
\end{equation}
Explicitly, for $\bm{K}_{\mathrm H}=(2\pi/a)(1,0,0)$ one finds $D\,H(\bm{k}+\bm{K}_{\mathrm H})\,D^{-1}=-H(\bm{k})$ with $D=\mathrm{diag}(1,-1,i,-i)$ in the basis of Eq.~(\ref{eq:HBloch}).
Equation~(\ref{eq:chiral}) maps the triple-Dirac cone at $\Gamma$ ($E=+\gamma$) onto that at H ($E=-\gamma$) with inverted spectrum, makes the density of states symmetric about $E=0$, and pins the Fermi level $E_F$ at half filling (one electron per site) at $E_F=0$.
Second, the weakly dispersing middle branch of the triple-Dirac cone is exactly flat, $E=+\gamma$, on the four $\{111\}$ planes through $\Gamma$ (for example, on the plane $\bm{k}\cdot(1,1,1)=0$), and hence also along the $\Gamma$--N lines, which lie in these planes.
Correspondingly, $E=-\gamma$ holds exactly on the $\{111\}$ planes through H.
Away from these planes the middle branch disperses quadratically. To leading order in $k$, $E=\gamma[1-(ka)^2/16]$ along $\langle100\rangle$ and $E=\gamma[1+(ka)^2/48]$ along $\langle111\rangle$.
This zero-field flat sheet is the origin of the nearly horizontal line at $E/\gamma=1$ in the $\bm{B}\parallel[111]$ spectra of Sect.~3.

Concerning the band filling, the four bands of the spinless model span $[-3\gamma,-\gamma]$, $[-\sqrt{3}\gamma,+\gamma]$, $[-\gamma,+\sqrt{3}\gamma]$, and $[+\gamma,+3\gamma]$, so that consecutive bands overlap in energy (bands 1 and 2, and bands 3 and 4, are degenerate at $E=\mp\sqrt{3}\gamma$ at the P points).
For an integer number of filled bands the Fermi level therefore does not coincide with a band edge.
For one, two, and three filled bands (1/2, 1, and 3/2 electrons per site when spin is included) we find $E_F\simeq-1.49\gamma$, $0$, and $+1.49\gamma$, respectively, and the system is metallic in all three cases.
The triple-Dirac points lie at the Fermi level for the fractional fillings $\nu=1/3$ (H point, $E_F=-\gamma$) and $\nu=2/3$ ($\Gamma$ point, $E_F=+\gamma$) of the four bands, i.e., for $2/3$ and $4/3$ electrons per site.
Numerically, the fraction of states below $E=+\gamma$ equals $2/3$ to within $2\times10^{-5}$.
At $E=\pm\gamma$ the flat sheets produce a van Hove-type enhancement of the density of states, so that the triple-Dirac cones are superimposed on a large background of states from other parts of the Brillouin zone.
\end{revblock}
\begin{figure*}[t]
    \centering
    \includegraphics[width=0.9\linewidth]{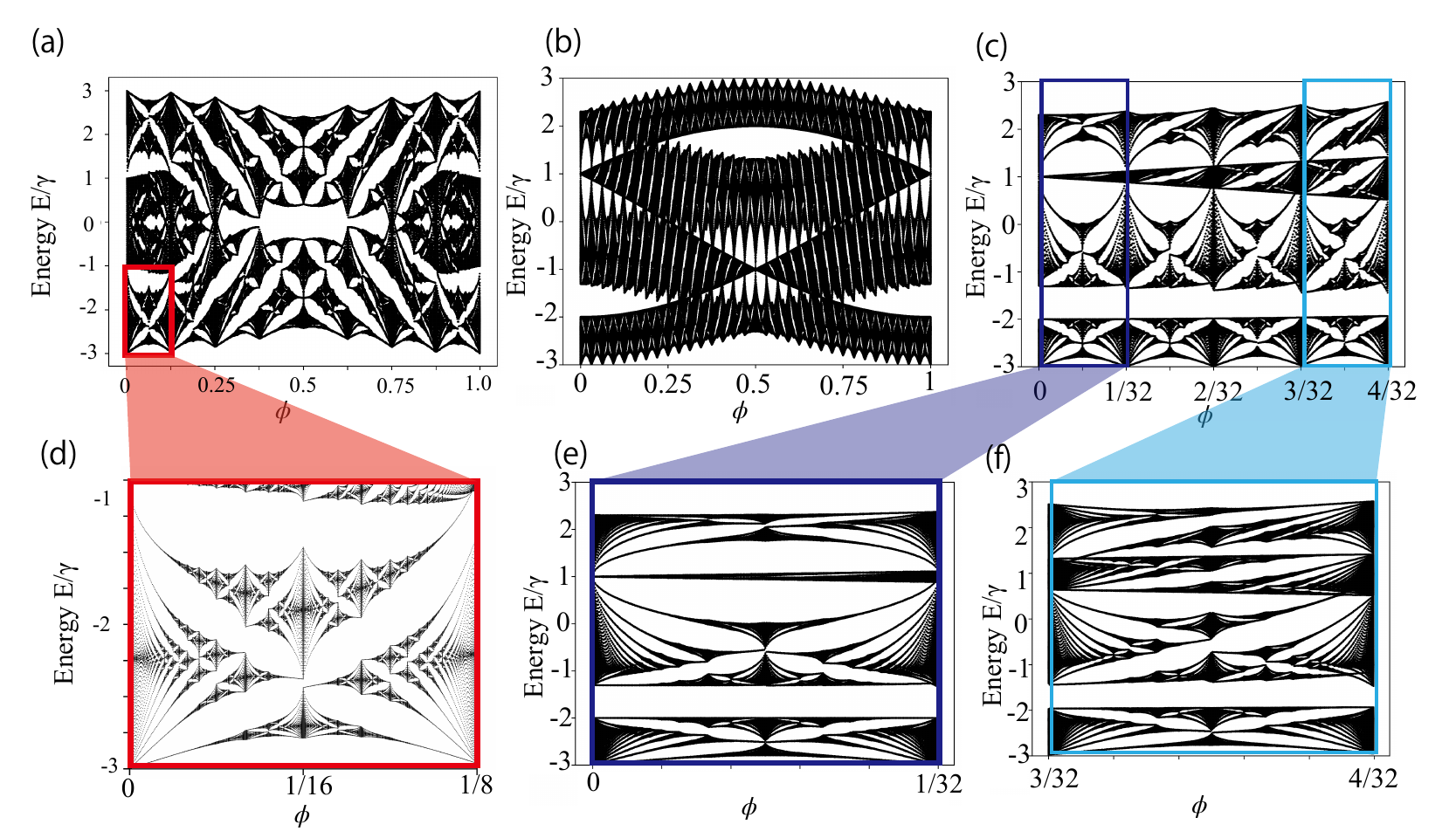}
    \caption{
        (Color online)
        Hofstadter energy spectra of the K$_4$ lattice at fixed field-parallel momentum $k_z = 0$ as functions of the dimensionless flux \rev{$\phi = BS_{\mathrm p}/\Phi_0$ through the elementary plaquette of the projected lattice, Eq.~(\ref{eq:flux_def}), where $S_{\mathrm p}=a^2/16$ for $\bm{B}\parallel[100]$ and $a^2/(32\sqrt{3})$ for $\bm{B}\parallel[111]$.
        The flux per projected unit cell is $8\phi$ and $32\phi$, respectively.
At $k_z=0$ the spectrum is even in $\phi$, and only $\phi\geq0$ is shown}.
        Energies are in units of the nearest-neighbor hopping amplitude $\gamma$, distinct from the energy--length product $\gamma\,a$ that enters the effective Hamiltonian in Sect.~4.
        (a) Full spectrum for $\bm{B}\parallel[100]$ \rev{over one period in $\phi$}. The box at $0<\phi<1/8$, $E/\gamma\leq-1$ marks the region enlarged in (d).
        (b) Full spectrum for $\bm{B}\parallel[111]$\rev{. The crossing at $E/\gamma=-1$ and $\phi=1/2$ is the H-point triple-Dirac cone, Eq.~(\ref{eq:half_shift})}.
        (c) Enlarged view of (b) over $0 < \phi < 4/32$. The boxes at $0<\phi<1/32$ and $3/32<\phi<4/32$ mark the regions further enlarged in (e) and (f).
        (d) Subpattern in the (100) spectrum on the scale $\phi = 1/8$ \rev{(one flux quantum per projected unit cell)}, reminiscent of the square-lattice Hofstadter butterfly.
        (e), (f) Subpatterns in the (111) spectrum on the scale $\phi = 1/32$ \rev{(one flux quantum per projected unit cell)}, reminiscent of the honeycomb- and triangular-lattice Hofstadter spectra, respectively\rev{.
        The nearly horizontal line at $E/\gamma=1$ in (e) and (f) descends from the exactly flat sheet of the $k_z=0$ slice at zero field (Sect.~2); at finite flux it consists of magnetic bands close to, but not exactly at, $E=\gamma$ (Sect.~3)}.
        The identification of each substructure with its two-dimensional counterpart relies on the flux periodicity and the hierarchy of miniband gaps rather than on an exact one-to-one mapping to those lattices.
    }
    \label{fig:k4_Peierls}
\end{figure*}

\section{Peierls Phase and Hofstadter Spectrum}

To incorporate a uniform magnetic field $\bm{B}$ into the tight-binding model, we apply the Peierls substitution.\cite{Peierls1933ZPhys}
The resulting Hamiltonian reads
\begin{equation}
H = -\gamma \sum_{\langle i,j \rangle} e^{-i\theta_{ij}}\, c_i^\dagger c_j + \mathrm{H.c.},
\label{eq:H_peierls}
\end{equation}
where $c_i^\dagger$ ($c_j$) creates (annihilates) an electron at site $i$ ($j$), $\langle i,j \rangle$ runs over the K$_4$ nearest-neighbor bonds defined in Sect.~2, and the Peierls phase associated with the bond from $j$ to $i$ is
\begin{equation}
\theta_{ij} = \frac{e}{\hbar c} \int_{\bm{R}_j}^{\bm{R}_i} \bm{A}(\bm{r}) \cdot d\bm{l},
\label{eq:peierls_phase}
\end{equation}
with $\bm{A}(\bm{r})$ the vector potential.
For lattice electrons in a uniform field, this construction generally produces a fractal energy spectrum, the Hofstadter butterfly being the canonical example for the square lattice.\cite{Harper1955,Brown1964PR,Zak1964PR,Hofstadter1976PRB,Thouless1982PRL,Kohmoto1985AnnPhys,Hasegawa1989PRL,Janecek2013PRB}

We consider two field orientations, $\bm{B}\parallel[100]$ and $\bm{B}\parallel[111]$.
For each case we rotate the laboratory frame so that the field is aligned with the new $\hat{\bm{z}}$ axis and adopt the Landau gauge
\begin{equation}
\bm{A}(\bm{r})=(0,\,Bx,\,0),
\qquad
\nabla\times\bm{A}=B\hat{\bm{z}}.
\label{eq:landau_gauge}
\end{equation}
Throughout this section, $(x,y,z)$ and $(k_x,k_y,k_z)$ denote coordinates and crystal momenta in this rotated frame\rev{, and $B$ is a signed quantity, positive when $\bm{B}$ points along $+\hat{\bm{z}}$}.
Because $\bm{A}$ depends only on $x$, translational symmetry along the field direction is preserved, so $k_z$ remains a good quantum number.
The three-dimensional problem then reduces, for each fixed $k_z$, to an effective two-dimensional magnetic problem on the lattice obtained by projecting K$_4$ onto the plane perpendicular to $\bm{B}$.

\begin{revblock}
We measure the magnetic field by the dimensionless flux through the elementary plaquette of the projected lattice,
\begin{equation}
\phi \equiv \frac{B S_{\mathrm p}}{\Phi_0},
\qquad
\Phi_0=\frac{hc}{e},
\label{eq:flux_def}
\end{equation}
where $S_{\mathrm p}$ is the area of the smallest closed loop of the projected lattice and $\Phi_0$ is the flux quantum.
The smallest loop is the four-membered ring of area $S_{\mathrm p}=a^2/16$ for $\bm{B}\parallel[100]$ [Fig.~\ref{fig:k4_structure}(d)] and the three-membered ring (triangle) of area $S_{\mathrm p}=a^2/(32\sqrt{3})$ for $\bm{B}\parallel[111]$ [Fig.~\ref{fig:k4_structure}(e)].
The projected Bravais lattice is a square lattice with unit-cell area $S=a^2/2=8S_{\mathrm p}$ for [100] and a triangular lattice with $S=a^2/\sqrt{3}=32S_{\mathrm p}$ for [111], so that the flux per projected unit cell is
\begin{equation}
\phi_{\mathrm{cell}}\equiv\frac{BS}{\Phi_0}=n_{\mathrm p}\,\phi,
\qquad
n_{\mathrm p}=
\begin{cases}
8 & (\bm{B}\parallel[100]),\\[2pt]
32 & (\bm{B}\parallel[111]).
\end{cases}
\label{eq:plaquette_flux}
\end{equation}
Each projected unit cell contains exactly two faces, the elementary plaquette and a complementary ring enclosing the remaining area $(n_{\mathrm p}-1)S_{\mathrm p}$.
The complementary ring is an eight-membered ring for [100] [Fig.~\ref{fig:k4_structure}(d)] and a nine-membered ring for [111] [Fig.~\ref{fig:k4_structure}(e)].
Since every closed loop of the projected lattice thus encloses an integer multiple of $S_{\mathrm p}$, the magnetic spectrum is exactly periodic in $\phi$ with period 1.
The values $\phi=1/8$ and $\phi=1/32$, which correspond to one flux quantum per projected unit cell, set the scales of the substructures discussed below.

For a rational $\phi_{\mathrm{cell}}=p/q$ with coprime integers $p$ and $q$, the Peierls phases are periodic up to magnetic translations, and the magnetic unit cell is enlarged by a factor of $q$ along one of the in-plane primitive directions perpendicular to $\bm{B}$.\cite{Brown1964PR,Zak1964PR}
\end{revblock}
We construct the tight-binding Hamiltonian on this enlarged unit cell, diagonalize it numerically for many values of the transverse Bloch momentum, and accumulate the eigenvalues to obtain the spectrum at fixed $k_z$ as a function of $\phi$.

The resulting Hofstadter spectra at $k_z = 0$ are summarized in Fig.~\ref{fig:k4_Peierls}.
Figures~\ref{fig:k4_Peierls}(a) and (b) display the full spectra for $\bm{B}\parallel[100]$ and $\bm{B}\parallel[111]$, respectively, and Fig.~\ref{fig:k4_Peierls}(c) shows an enlarged view of the low-flux region $0 < \phi < 4/32$ of panel~(b).
For $\bm{B}\parallel[100]$, a substructure resembling the Hofstadter butterfly of the square lattice appears on the scale $\phi=1/8$ [Fig.~\ref{fig:k4_Peierls}(d)].
\rev{This scale corresponds to one flux quantum per projected unit cell, $\phi_{\mathrm{cell}}=1$.
Within each interval $m<\phi_{\mathrm{cell}}<m+1$ ($m$ integer) the magnetic unit cell runs through all sizes $q$, as in the Hofstadter problem of the square Bravais lattice of the (100) projection, whereas the elementary four-membered plaquette of area $S/8$ sets the overall period $\phi=1$ of the spectrum.}
For $\bm{B}\parallel[111]$, two distinct substructures emerge on the finer scale $\phi=1/32$\rev{, again one flux quantum per (triangular) projected unit cell}.
One is reminiscent of the honeycomb-lattice butterfly [Fig.~\ref{fig:k4_Peierls}(e)] and the other of the triangular-lattice butterfly [Fig.~\ref{fig:k4_Peierls}(f)], in agreement with the honeycomb-plus-triangular nature of the (111) projection [Fig.~\ref{fig:k4_structure}(e)].
The simultaneous appearance of both subpatterns in the same spectrum is a direct signature of the dual character of the (111) projection.
We emphasize that the labels ``square-like'', ``honeycomb-like'', and ``triangular-like'' refer to the flux periodicity and the hierarchy of miniband gaps of the substructures rather than to an exact one-to-one mapping of the K$_4$ magnetic spectrum onto the spectra of those two-dimensional lattices.
Quantitative agreement with the projected plaquette geometry is established here through the flux scales of Eq.~(\ref{eq:plaquette_flux}).

\begin{revblock}
The alternation between the honeycomb-like and triangular-like subpatterns with increasing flux [Fig.~\ref{fig:k4_Peierls}(c)] can be traced to the two plaquettes of the (111) projection.
At the boundaries $\phi=m/32$ of the flux windows the flux per projected unit cell is an integer, the magnetic unit cell coincides with the projected unit cell, and the spectrum collapses to four bands.
The Peierls phases, however, do not disappear at these values but form a staggered pattern, with the phase $2\pi m/32$ around each elementary triangle and $-2\pi m/32$ (mod $2\pi$) around each nine-membered ring.
The spectrum in the $m$-th window therefore interpolates, through the fractional part of $\phi_{\mathrm{cell}}$, between four-band structures that themselves evolve with $m$.
For $m=0$ the four bands at $k_z=0$ span $[-3,-2]$, $[-1.30,1]$, $\{1\}$, and $[1,2.30]$ (in units of $\gamma$); the flat third band at $m=0$ is the zero-field flat sheet of Sect.~2, which the other bands touch at $\Gamma$.
For $m=4$ they have become $[-2.95,-1.93]$, $[-1.49,0.52]$, $[0.52,1.41]$, and $[1.41,2.58]$, and at $m=8$ ($\phi=1/4$) the spectrum is symmetric about $E=0$.
In the first window the small triangles are essentially transparent to the field ($\phi<1/32$), and the spectrum is governed by the honeycomb-like network formed by the nine-membered rings, which produces the honeycomb-like pattern with the nearly flat level at $E/\gamma=1$ [Fig.~\ref{fig:k4_Peierls}(e)].
It is exactly flat only at $\phi=0$.
At finite flux $\phi_{\mathrm{cell}}=p/q$ it splits into $q$ magnetic bands (the central $q$ of the $4q$ bands), whose deviation from $E=\gamma$ grows with the flux, from about $0.02\gamma$ at $|\phi|=1/200$ to about $0.1\gamma$ at $\phi=1/32$.
With increasing $m$ the phase accumulated around the triangles progressively broadens the flat band, removes the Dirac-like fans at $E/\gamma=1$, and yields the triangular-like pattern [Fig.~\ref{fig:k4_Peierls}(f)].
The relevant geometric parameter is thus the area ratio $S_{\mathrm p}/S=1/32$ between the triangular sublattice motif and the projected unit cell, which fixes both the window width $1/32$ and the staggered phase $m/32$ in the $m$-th window.

The spectra at finite $k_z$, collected in the Appendix, are related to the $k_z=0$ spectra by an exact relation.
Because of the chirality of the K$_4$ network, the elementary plaquette of the projected lattice is not a closed loop in three dimensions but the projection of one turn of a helix along the field.
The helix is generated by a four-fold screw axis for [100] and by a three-fold screw axis for [111], and the two helices are of opposite handedness.
One turn advances by the lattice period $c$ along $\bm{B}$, with $c=a$ for [100] and $c=a\sqrt{3}/2$ for [111].
In the reduced two-dimensional problem the Bloch factor $e^{\pm ik_zc}$ accumulated along one turn therefore adds to the Peierls phase of the elementary plaquette, $2\pi\phi\to2\pi(\phi\pm\kappa)$, with
\begin{equation}
\kappa\equiv\frac{k_z c}{2\pi},
\label{eq:kappa}
\end{equation}
whereas the complementary ring, which winds in the opposite sense, acquires $\mp2\pi\kappa$.
The field-parallel momentum thus acts as an additional flux through the elementary plaquettes, compensated by the opposite flux through the complementary rings, with no net flux per unit cell.
Whenever $n_{\mathrm p}\kappa$ is an integer, the phase $\mp2\pi\kappa$ of the complementary ring coincides modulo $2\pi$ with the phase $\pm2\pi(n_{\mathrm p}-1)\kappa$ that a uniform shift of $\phi$ by $\pm\kappa$ would produce, and the finite-$k_z$ spectrum is exactly a rigid shift of the $k_z=0$ spectrum along the flux axis,
\begin{equation}
E(k_z,\phi)=E(0,\phi\pm\kappa)
\qquad\text{for}\quad
k_z=\frac{2\pi n}{n_{\mathrm p}\,c}\quad(n\ \text{integer}).
\label{eq:kz_shift}
\end{equation}
The sign in Eq.~(\ref{eq:kz_shift}) is fixed by the handedness of the elementary helix and is therefore opposite for the two field orientations: with $\phi$, $k_z$, and $\kappa$ signed as in Eqs.~(\ref{eq:landau_gauge}), (\ref{eq:flux_def}), and~(\ref{eq:kappa}), the lower sign applies to $\bm{B}\parallel[100]$ and the upper sign to $\bm{B}\parallel[111]$.
This assignment holds for either sign of $B$, because time-reversal symmetry gives $E(k_z,\phi)=E(-k_z,-\phi)$, under which Eq.~(\ref{eq:kz_shift}) is invariant.
We have verified Eq.~(\ref{eq:kz_shift}) numerically to machine precision for both signs of $B$.
For intermediate $k_z$ the spectrum deforms continuously between the shifted patterns (Appendix).
For $\bm{B}\parallel[100]$ the shift is $n/8$, i.e., the width of one square-like substructure for $k_za=\pi/4$; for $\bm{B}\parallel[111]$ it is $n/32$, the width of one fine substructure for $k_za=\pi/(8\sqrt{3})$.
The spectra of Fig.~\ref{fig:finite_kz}, and the lattice Landau bands of Sect.~\ref{sec:scaleconnect}, are computed for $B<0$ in the convention of Eqs.~(\ref{eq:H_peierls})--(\ref{eq:landau_gauge}).
In Fig.~\ref{fig:finite_kz} the horizontal axis is the magnitude of the flux, as in Fig.~\ref{fig:k4_Peierls}; there the $k_z=0$ pattern of Fig.~\ref{fig:k4_Peierls} appears shifted toward lower $\phi$ by $\kappa$ for [100] and toward higher $\phi$ for [111].

Combining Eq.~(\ref{eq:kz_shift}) with the chiral symmetry (\ref{eq:chiral}) locates the second triple-Dirac cone in the magnetic spectra.
Since the sublattice signs in $D$ and the shift by $\bm{K}_{\mathrm H}$ leave the Peierls phases unchanged, the chiral symmetry (\ref{eq:chiral}) persists in the magnetic problem.
For $\bm{B}\parallel[100]$ the H points $(2\pi/a)(0,\pm1,0)$ and $(2\pi/a)(0,0,\pm1)$ lie in the plane $k_z=0$ together with $\Gamma$, and $\bm{K}_{\mathrm H}=(2\pi/a)(1,0,0)$ is parallel to $\bm{B}$ with $K_{\mathrm H}c=2\pi$, so that the momentum shift in Eq.~(\ref{eq:chiral}) leaves $k_z$ unchanged.
The (100) spectrum is therefore symmetric about $E=0$ for every $k_z$ and $\phi$, and at $k_z=0$ the Landau fans of the $\Gamma$ and H cones appear simultaneously at $E=\pm\gamma$ [Figs.~\ref{fig:k4_Peierls}(a) and (d)].
For $\bm{B}\parallel[111]$ all H points have $k_z=\pm2\pi/(\sqrt{3}a)$, i.e., $\kappa=\pm1/2$, so that Eqs.~(\ref{eq:chiral}) and~(\ref{eq:kz_shift}) give
\begin{equation}
E(k_z,\phi)=-E\!\left(k_z,\phi+\tfrac{1}{2}\right)
\qquad(\bm{B}\parallel[111]).
\label{eq:half_shift}
\end{equation}
The (111) spectrum at $k_z=0$ therefore contains the H-point cone as well.
The crossing at $E/\gamma=-1$ and $\phi=1/2$ in Fig.~\ref{fig:k4_Peierls}(b) is the mirror image of the $\Gamma$-point cone at $\phi=0$, and the nearly horizontal line at $E/\gamma=1$ is accompanied by one at $E/\gamma=-1$ around $\phi=1/2$.
Equivalently, at $k_z=2\pi/(\sqrt{3}a)$ the H cone appears at $\phi=0$ [Fig.~\ref{fig:finite_kz}(f)].
\end{revblock}

These results demonstrate that, for each $k_z$, the K$_4$ lattice exhibits a Hofstadter spectrum whose self-similar substructures faithfully report the projected lattice geometry\rev{, and that the field-parallel momentum and the chirality of the network enter the spectrum through the exact relations (\ref{eq:kz_shift}) and (\ref{eq:half_shift})}.
The connectivity of the K$_4$ network is thereby quantitatively imprinted on its magnetic energy spectrum.

\section{Weak-Field Limit and Landau Levels}

To account analytically for the low-flux structure of the Hofstadter spectrum, we now derive the Landau-level spectrum from the low-energy effective Hamiltonian around the triple-Dirac point at $\Gamma$.
As in Sect.~3, the field is taken along the rotated $\hat{\bm{z}}$ axis, so that $k_z$ denotes the Bloch momentum parallel to $\bm{B}$ and $(k_x,k_y)$ the transverse components.

Near $\Gamma$, the low-energy electronic structure of the K$_4$ lattice is captured by the three-component effective Hamiltonian\cite{Tsuchiizu2016PRB}
\begin{equation}
H_{\mathrm{eff}} = \frac{\gamma\,a}{2}
\begin{pmatrix}
    k_z & \frac{1}{\sqrt{2}} k_{-} & 0 \\[2pt]
    \frac{1}{\sqrt{2}} k_{+} & 0 & \frac{1}{\sqrt{2}} k_{-} \\[2pt]
    0 & \frac{1}{\sqrt{2}} k_{+} & -k_z
\end{pmatrix},
\qquad
k_{\pm} = k_x \pm i k_y .
\label{eq:Heff}
\end{equation}
Here the matrix entries are components of the wavevector $\bm{k}$ measured from $\Gamma$, and the energy--length product $\gamma\,a$ serves as the overall coefficient that gives the matrix the dimensions of energy.
\rev{Equation~(\ref{eq:Heff}) is the first-order $\bm{k}\cdot\bm{p}$ Hamiltonian of the threefold degenerate level at $\Gamma$.
Expanding Eq.~(\ref{eq:HBloch}) to linear order in $\bm{k}$ and projecting onto the three eigenvectors of $H(\bm{0})$ with eigenvalue $+\gamma$ gives, apart from the constant $+\gamma$, $H_{\mathrm{eff}}=(\gamma a/2)\,\bm{k}\cdot\bm{J}$, where $\bm{J}=(J_x,J_y,J_z)$ are spin-1 matrices obeying $[J_x,J_y]=iJ_z$ and $\bm{J}^2=2$.
We have verified this numerically, including the value $\gamma a/2$ of the prefactor.
Equation~(\ref{eq:Heff}) is this Hamiltonian written in the basis of $J_z$ eigenstates, and because $\bm{k}\cdot\bm{J}$ is rotationally covariant the same form holds in the rotated frame of Sect.~3 for either field orientation.}
The corresponding eigenvalues $0$ and $\pm (\gamma\,a)|\bm{k}|/2$ reproduce the triple-Dirac cone of Fig.~\ref{fig:k4_structure}(c) with Fermi velocity $\hbar v_F = \gamma\,a/2$\rev{, isotropic to this order}.
The construction of Landau levels from such a Hamiltonian follows the standard minimal-coupling procedure that has been developed extensively for graphite, graphene, and related low-pseudospin systems.\cite{Wallace1947PR,McClure1956PR,McClure1957PR,Slonczewski1958PR,McClure1960PR,Dresselhaus1974PRB,Nakao1976JPSJ,McCann2006PRL,CastroNeto2009RMP,Goerbig2011RMP,McCann2013RPP,Raoux2014PRL,Malcolm2014PRB,Illes2016PRB}

Applying the Landau gauge $\bm{A}=(0,Bx,0)$ and the minimal-coupling substitution $\bm{k}\to\bm{\Pi}=\bm{k}-(e/\hbar c)\bm{A}$, the kinetic momentum components become
\begin{equation}
\Pi_x = -i \partial_x, \quad
\Pi_y = -i \partial_y - \frac{eB}{\hbar c}\,x, \quad
\Pi_z = -i \partial_z .
\label{eq:Pi_xyz}
\end{equation}
\rev{Equation~(\ref{eq:Pi_xyz}) corresponds to a carrier of charge $+e$, whereas the Peierls phase of Eqs.~(\ref{eq:H_peierls}) and~(\ref{eq:peierls_phase}) corresponds to the electron charge $-e$.
The two conventions are related by $B\to-B$, which in all results below amounts to $k_z\to-k_z$.}
With Eq.~(\ref{eq:Pi_xyz}), Eq.~(\ref{eq:Heff}) becomes
\begin{equation}
\hat{H} = \gamma\,a
\begin{pmatrix}
    \frac{1}{2} \Pi_z & \frac{1}{2\sqrt{2}} \Pi_{-} & 0 \\[2pt]
    \frac{1}{2\sqrt{2}} \Pi_{+} & 0 & \frac{1}{2\sqrt{2}} \Pi_{-} \\[2pt]
    0 & \frac{1}{2\sqrt{2}} \Pi_{+} & -\frac{1}{2} \Pi_z
\end{pmatrix},
\qquad
\Pi_\pm = \Pi_x \pm i\Pi_y.
\label{eq:Hhat}
\end{equation}
Since $\gamma\,a$ is an overall multiplicative factor in Eq.~(\ref{eq:Hhat}), the physical eigenvalue equation $\hat{H}\Psi = E\Psi$ reduces, after dividing both sides by $\gamma\,a$, to a spectral problem for the rescaled eigenvalue
\begin{equation}
\varepsilon \equiv \frac{E}{\gamma\,a},
\label{eq:E_rescaled}
\end{equation}
which has dimensions of inverse length.
Throughout Sects.~\ref{sec:LL-method1} and~\ref{sec:LL-method2}, $\varepsilon$ denotes this rescaled eigenvalue; \rev{here $E$ is measured from the cone center, and the total energy, $\gamma+(\gamma a)\varepsilon$ for the $\Gamma$ cone, is restored} in Sect.~\ref{sec:scaleconnect}.
A direct calculation gives the bosonic commutation relation
\begin{equation}
[\Pi_+,\Pi_-]=s,
\qquad
s\equiv\frac{2eB}{\hbar c}.
\label{eq:comm_s}
\end{equation}
\rev{Here $B>0$ is assumed in the convention of Eq.~(\ref{eq:Pi_xyz}), so that $s>0$; for $B<0$ the roles of $\Pi_\pm$ are interchanged, equivalent to $k_z\to-k_z$.}
We therefore introduce ladder operators (denoted with a hat to distinguish them from the lattice constant $a$)
\begin{equation}
\hat{a}=\frac{1}{\sqrt{s}}\,\Pi_+,
\qquad
\hat{a}^\dagger=\frac{1}{\sqrt{s}}\,\Pi_-,
\qquad [\hat{a},\hat{a}^\dagger]=1,
\label{eq:ladder_def}
\end{equation}
and denote the eigenstates of $\hat{a}^\dagger \hat{a}$ by $\{|n\rangle\}$ with $n=0,1,2,\dots$.
These satisfy
\begin{align}
\Pi_+ |n\rangle &= \sqrt{n s}\,|n-1\rangle, \label{eq:PiPlus_action}\\
\Pi_- |n\rangle &= \sqrt{(n+1)s}\,|n+1\rangle, \label{eq:PiMinus_action}\\
\Pi_-\Pi_+ &= s\,n,
\qquad
\Pi_+\Pi_- = s\,(n+1).
\label{eq:PiPi_id}
\end{align}
In what follows we derive the Landau-level spectrum in two equivalent ways: by direct reduction of the coupled differential equations (Sect.~4.1) and by projection onto the oscillator basis (Sect.~4.2).

\subsection{Reduction of the Coupled Equations}
\label{sec:LL-method1}

Assuming plane-wave dependence $e^{ik_z z}$ along the field direction, we write the three-component spinor as $\Psi = (\Psi_1, \Psi_2, \Psi_3)^{\mathrm T}$.
After dividing both sides of $\hat{H}\Psi=E\Psi$ by $\gamma\,a$ (cf.\ Eq.~(\ref{eq:E_rescaled})), the eigenvalue equation for the rescaled eigenvalue $\varepsilon$ reads
\begin{align}
& \tfrac{1}{2} k_z \Psi_1 + \tfrac{1}{2\sqrt{2}} \Pi_{-} \Psi_2 = \varepsilon \Psi_1, \label{eq:coupled1}\\
& \tfrac{1}{2\sqrt{2}} \Pi_{+} \Psi_1 + \tfrac{1}{2\sqrt{2}} \Pi_{-} \Psi_3 = \varepsilon \Psi_2, \label{eq:coupled2}\\
& \tfrac{1}{2\sqrt{2}} \Pi_{+} \Psi_2 - \tfrac{1}{2} k_z \Psi_3 = \varepsilon \Psi_3. \label{eq:coupled3}
\end{align}
Eliminating $\Psi_1$ and $\Psi_3$ from Eqs.~(\ref{eq:coupled1})--(\ref{eq:coupled3}) yields a single equation for $\Psi_2$,
\begin{equation}
\begin{split}
\Bigl[ \tfrac{1}{8}(\varepsilon + \tfrac{1}{2} k_z)(\Pi_{-}\Pi_{+} + s) \\
+ \tfrac{1}{8}(\varepsilon - \tfrac{1}{2} k_z)\Pi_{-}\Pi_{+} \Bigr] \Psi_2
= \varepsilon\!\left(\varepsilon^2 - \tfrac{k_z^2}{4}\right)\Psi_2,
\end{split}
\label{eq:psi2eq}
\end{equation}
where we have used $\Pi_+\Pi_- = \Pi_-\Pi_+ + s$ from Eq.~(\ref{eq:comm_s}).
Substituting $\Pi_-\Pi_+ = s\,n$ from Eq.~(\ref{eq:PiPi_id}) and rearranging gives the cubic secular equation
\begin{equation}
\varepsilon^3 - \frac{s}{4}\left(n+\frac{1}{2}+\frac{k_z^2}{s}\right)\varepsilon - \frac{s}{16}\,k_z = 0,
\label{eq:cubicLL}
\end{equation}
\rev{The elimination assumes $\varepsilon\neq\pm k_z/2$, since Eqs.~(\ref{eq:coupled1}) and~(\ref{eq:coupled3}) were divided by $\varepsilon\mp k_z/2$ and the resulting denominators were cleared by multiplying by $\varepsilon^2-k_z^2/4$ to obtain Eq.~(\ref{eq:psi2eq}).
Eigenstates at these exceptional energies must be examined directly in Eqs.~(\ref{eq:coupled1})--(\ref{eq:coupled3}).}
For $n\geq 1$ the three roots of Eq.~(\ref{eq:cubicLL}) yield the rescaled Landau-level eigenvalues $\varepsilon$ of the triple-Dirac Hamiltonian, with the corresponding energies \rev{measured from the cone center} recovered as \rev{$\delta E = (\gamma\,a)\,\varepsilon$}.
The exceptional $n=0$ sector requires separate treatment: \rev{one factor of the cubic corresponds to no eigenstate, and an anomalous zero mode at the exceptional energy $\varepsilon=k_z/2$ must be added}.
This sector is analyzed in Sect.~\ref{sec:LL-method2}.

\subsection{Oscillator-Basis Derivation and the $n=0$ Sector}
\label{sec:LL-method2}

The same secular equation can be obtained algebraically by projecting $\hat{H}$ onto a finite oscillator-basis sector.
For $n\geq 1$ we take the ansatz
\begin{equation}
\Psi^{(n)} =
\begin{pmatrix}
c_1 |n+1\rangle \\
c_2 |n\rangle \\
c_3 |n-1\rangle
\end{pmatrix},
\label{eq:basis_choice}
\end{equation}
which is closed under the action of $\Pi_\pm$ on each component.
In this basis $\hat{H}$ is represented by the $3\times 3$ matrix
\begin{equation}
D_n =
\frac{1}{2}\begin{pmatrix}
 k_z & \sqrt{(n+1)s/2} & 0 \\[2pt]
 \sqrt{(n+1)s/2} & 0 & \sqrt{ns/2} \\[2pt]
0 & \sqrt{ns/2} & -k_z
\end{pmatrix}.
\label{eq:Dn}
\end{equation}
Expanding the secular determinant along the first row gives
\begin{equation}
\det(D_n - \varepsilon I)
= -\varepsilon^3 + \frac{1}{4}\!\left(k_z^2 + \frac{(2n+1)s}{2}\right)\!\varepsilon + \frac{s\,k_z}{16},
\label{eq:DnDet}
\end{equation}
so that $\det(D_n - \varepsilon I)=0$ reproduces the cubic of Eq.~(\ref{eq:cubicLL}) without further manipulation, after rewriting the bracket as $(s/4)(n+1/2+k_z^2/s)$.

\begin{figure*}[t]
\centering
\includegraphics[width=\textwidth]{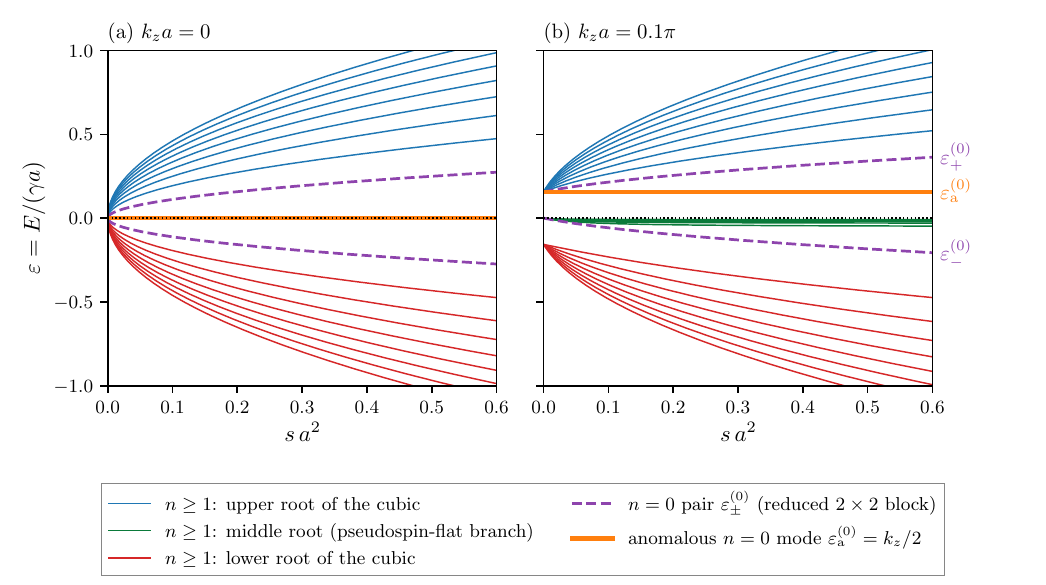}
\caption{
(Color online)
Landau-level spectrum of the K$_4$ triple-Dirac cone obtained from the effective Hamiltonian Eq.~(\ref{eq:Heff}), for (a) $k_z a = 0$ and (b) $k_z a = 0.1\pi$.
The horizontal axis is $s\,a^2$, with $s = 2eB/(\hbar c)$ the field strength of Eq.~(\ref{eq:comm_s}), and the vertical axis is the rescaled eigenvalue $\varepsilon = E/(\gamma\,a)$ of Eq.~(\ref{eq:E_rescaled}).
\rev{Thin solid lines: the three roots of the cubic Eq.~(\ref{eq:cubicLL}) for $n=1,\dots,8$, for the upper, middle, and lower branches in order of decreasing $\varepsilon$ (online: blue, green, and red).
Dashed lines: the $n=0$ pair $\varepsilon^{(0)}_\pm$ of Eq.~(\ref{eq:n0_quadratic}).
Thick solid line: the single anomalous $n=0$ level $\varepsilon^{(0)}_{\mathrm a}=k_z/2$ of Eq.~(\ref{eq:anomalous}).
At $k_z=0$ (a) all middle roots and the anomalous level coincide at $\varepsilon=0$, and the remaining levels form the relativistic fan $\varepsilon=\pm\tfrac{1}{2}\sqrt{s(n+\tfrac{1}{2})}$.
At finite $k_z$ (b) the middle roots are slightly displaced from $\varepsilon=0$, the $n=0$ pair splits off from the fan, and the anomalous level is pinned at $\varepsilon=k_z/2$.
The relation between $s\,a^2$ and the flux $\phi$ of Fig.~\ref{fig:k4_Peierls} is Eq.~(\ref{eq:s_phi}).}
\rev{The relation between $\varepsilon$ and the energies of Figs.~\ref{fig:k4_structure}(c) and~\ref{fig:k4_Peierls} is given in Sect.~\ref{sec:scaleconnect}.}
}
\label{fig:landau}
\end{figure*}

The $n=0$ sector requires separate treatment because the third component $|n-1\rangle = |-1\rangle$ in Eq.~(\ref{eq:basis_choice}) is absent.
Setting $c_3 = 0$ and removing the third row and column of Eq.~(\ref{eq:Dn}) leaves the $2\times 2$ block
\begin{equation}
\widetilde{D}_0 = \frac{1}{2}\begin{pmatrix} k_z & \sqrt{s/2} \\[2pt] \sqrt{s/2} & 0 \end{pmatrix},
\end{equation}
with eigenvalues
\begin{equation}
\varepsilon^{(0)}_{\pm} = \frac{k_z}{4} \pm \frac{1}{4}\sqrt{k_z^2 + 2s}.
\label{eq:n0_quadratic}
\end{equation}
In addition, because $\Pi_+|0\rangle = 0$, the configuration
$\Psi^{(0)}_{\mathrm{a}} = (|0\rangle,\,0,\,0)^{\mathrm T}$
is itself an eigenstate of $\hat{H}$ with rescaled eigenvalue
\begin{equation}
\varepsilon^{(0)}_{\mathrm{a}} = \frac{k_z}{2} ,
\label{eq:anomalous}
\end{equation}
analogous to the Dirac zero mode of the lowest Landau level in monolayer graphene.\cite{CastroNeto2009RMP,Goerbig2011RMP}
Equations~(\ref{eq:n0_quadratic}) and~(\ref{eq:anomalous}) together provide the three Landau-level eigenvalues of the $n=0$ multiplet.
\rev{In the approach of Sect.~4.1 the $n=0$ cubic~(\ref{eq:cubicLL}) factorizes as $(\varepsilon+k_z/2)(\varepsilon^2-k_z\varepsilon/2-s/8)=0$, and the quadratic factor reproduces the pair~(\ref{eq:n0_quadratic}).
The linear factor is an artifact of the multiplication by $\varepsilon+k_z/2$, which is superfluous for $n=0$ because $\Pi_-\Pi_+|0\rangle=0$.
It corresponds to no additional eigenstate, although its value may coincide with a physical level at isolated parameters (with $\varepsilon^{(0)}_{\mp}$ at $k_z=\pm\sqrt{s}/2$).
The anomalous mode~(\ref{eq:anomalous}) lies at the exceptional energy $\varepsilon=k_z/2$ and follows directly from Eqs.~(\ref{eq:coupled1})--(\ref{eq:coupled3}).
For $k_z\neq0$, Eq.~(\ref{eq:coupled1}) forces $\Psi_2=0$, Eq.~(\ref{eq:coupled3}) then gives $\Psi_3=0$, and Eq.~(\ref{eq:coupled2}) requires $\Pi_+\Psi_1=0$, so that $\Psi^{(0)}_{\mathrm a}$ is the only solution.
We emphasize that the term ``anomalous'' refers only to this single state, which lies outside the blocks of Eq.~(\ref{eq:basis_choice}) and $\widetilde{D}_0$.
The condition $\Psi_2=0$ alone does not single it out, since at $k_z=0$ the middle root $\varepsilon=0$ of Eq.~(\ref{eq:cubicLL}) has the eigenvector $(\sqrt{n}\,|n+1\rangle,0,-\sqrt{n+1}\,|n-1\rangle)^{\mathrm T}/\sqrt{2n+1}$ with $\Psi_2=0$ for every $n\geq1$.
The pair $\varepsilon^{(0)}_\pm$ of Eq.~(\ref{eq:n0_quadratic}) are ordinary $n=0$ levels of the reduced $2\times2$ block, and for $n\geq1$ all three roots of Eq.~(\ref{eq:cubicLL}) are ordinary levels.
Figure~\ref{fig:landau} distinguishes these three groups explicitly.}

The Landau-level spectrum derived above is shown in Fig.~\ref{fig:landau}.
At $k_z = 0$ [Fig.~\ref{fig:landau}(a)] the secular equation factorizes as $\varepsilon[\varepsilon^2 - (s/4)(n+\tfrac{1}{2})] = 0$, giving a flat branch at $\varepsilon=0$ plus a relativistic pair $\varepsilon = \pm\tfrac{1}{2}\sqrt{s(n+\tfrac{1}{2})}$ for every $n$.
In this limit, the anomalous mode of Eq.~(\ref{eq:anomalous}) merges with the flat branch.
The flat branch is a direct consequence of the pseudospin-flat band of the effective Hamiltonian and is reminiscent of the central zero-energy levels in $\alpha$-$T_3$ and other higher-pseudospin lattice models with flat bands of geometric origin.\cite{Sutherland1986PRB,Raoux2014PRL,Malcolm2014PRB,Illes2016PRB}
For finite $k_z$ [Fig.~\ref{fig:landau}(b)], the three branches couple via the diagonal $\pm k_z/2$ entries of Eq.~(\ref{eq:Heff}).
\rev{The middle roots are slightly displaced from $\varepsilon=0$ and the $n=0$ pair splits off from the relativistic fan,} while the anomalous $n=0$ state remains pinned at $\varepsilon^{(0)}_{\mathrm{a}} = k_z/2$.

\begin{revblock}
It is instructive to relate Fig.~\ref{fig:landau}(b) to the zero-field dispersion at fixed $k_z$.
For $\bm{k}$ displaced from $\Gamma$ along the field, Eq.~(\ref{eq:Heff}) gives, as a function of the transverse momentum $k_\perp=\sqrt{k_x^2+k_y^2}$,
\begin{equation}
\varepsilon=0,
\qquad
\varepsilon=\pm\tfrac{1}{2}\sqrt{k_\perp^2+k_z^2}.
\label{eq:hyperbolic}
\end{equation}
The two linear branches thus become hyperbolic, with a gap $\gamma a|k_z|$ at $k_\perp=0$, while the pseudospin-flat branch remains exactly at $\varepsilon=0$ for every $k_z$.
On the lattice, the latter acquires only the weak, direction-dependent quadratic dispersion quoted in Sect.~2, e.g., $E-\gamma=-\gamma(k_za)^2/16$ for $\bm{k}$ along [100].
The $n\geq1$ levels of Fig.~\ref{fig:landau}(b) are the magnetic quantization of these branches.
The upper and lower roots of Eq.~(\ref{eq:cubicLL}) behave as $\pm\tfrac{1}{2}\sqrt{k_z^2+s(n+\tfrac{1}{2})}$ to leading order, like the Landau levels of a massive Dirac fermion with a mass proportional to $k_z$, whereas the middle roots are displaced from the flat branch only through the interband (pseudospin) coupling, $\varepsilon\simeq-s\,k_z/[4(k_z^2+s(n+\tfrac{1}{2}))]$, and return to $\varepsilon=0$ as $s\to0$.
The anomalous level $\varepsilon^{(0)}_{\mathrm a}=k_z/2$ is special in this respect.
For $k_z>0$ it sits exactly at the bottom of the upper hyperbolic branch, $\varepsilon=+|k_z|/2$ at $k_\perp=0$, for $k_z<0$ at the top of the lower one, and it disperses linearly through the flat branch as $k_z$ changes sign.
It is therefore the spin-1 counterpart of the chiral zeroth Landau level of a Weyl node,\cite{Armitage2018RMP} its direction of propagation being fixed by the product of the chirality of the node and the sign of $B$ (reversing $\bm{B}$ maps $k_z\to-k_z$ in Eqs.~(\ref{eq:cubicLL})--(\ref{eq:anomalous})).
The H-point cone, related to the $\Gamma$ cone by Eq.~(\ref{eq:chiral}), carries the counter-propagating partner.
\end{revblock}

\begin{figure*}[t]
\centering
\includegraphics[width=\textwidth,clip]{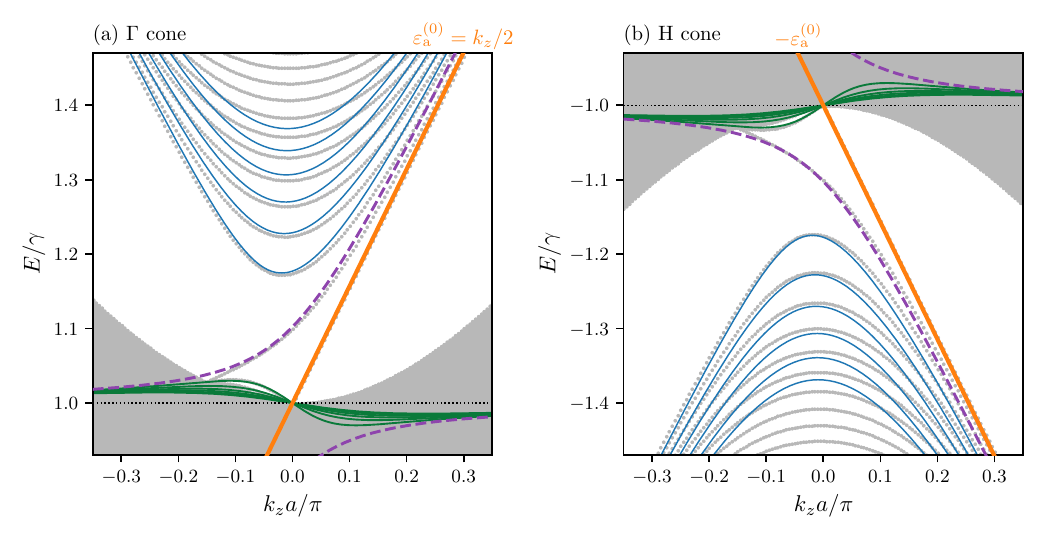}
\caption{
(Color online)
Landau bands $E(k_z)$ of the K$_4$ lattice for $\bm{B}\parallel[100]$ at the fixed flux $|\phi|=1/2400$ ($|\phi_{\mathrm{cell}}|=1/300$, $s\,a^2=8\pi/300\simeq0.084$), around (a) the $\Gamma$ cone and (b) the H cone.
Dots: eigenvalues of the magnetic tight-binding Hamiltonian of Sect.~3 for one transverse Bloch momentum (the levels are dispersionless in the transverse momenta at this flux).
The shaded regions are the continua of the other bands in the same $k_z$ slice.
Lines: Landau levels of the effective Hamiltonian, $E/\gamma=\pm(1+\varepsilon)$ with $\varepsilon$ from Eq.~(\ref{eq:cubicLL}) for $n=1$--$6$ (thin solid lines, the upper and middle roots; the lower roots lie outside the displayed energy range), from Eq.~(\ref{eq:n0_quadratic}) for the $n=0$ pair (dashed), and from Eq.~(\ref{eq:anomalous}) for the anomalous level (thick solid), using Eq.~(\ref{eq:s_phi}).
The lattice spectrum is computed for $B<0$ in the convention of Eqs.~(\ref{eq:H_peierls})--(\ref{eq:landau_gauge}) [$B>0$ in that of Eq.~(\ref{eq:Pi_xyz})], so that the anomalous level of the $\Gamma$ cone disperses upward as in Eq.~(\ref{eq:anomalous}); for the opposite sign of $B$ the spectrum is mirrored, $k_z\to-k_z$.
}
\label{fig:landau_bands}
\end{figure*}

\subsection{\rev{Relation to the Lattice Hofstadter Spectrum}}
\label{sec:scaleconnect}

\begin{revblock}
Although Eq.~(\ref{eq:Heff}) is obtained by expanding the tight-binding Hamiltonian around $\Gamma$, the variables of the continuum Landau-level problem still have to be related to the dimensionless energy and flux of the lattice Hofstadter calculation of Sect.~3 before the two can be compared.
The physical energies follow from Eq.~(\ref{eq:E_rescaled}) as $E=\gamma+(\gamma a)\varepsilon$ for the $\Gamma$ cone and, by the chiral symmetry (\ref{eq:chiral}), as $E=-\gamma-(\gamma a)\varepsilon$ for the H cone.
With energies in units of $\gamma$ and lengths in units of $a$, the plotted $\varepsilon$ of Fig.~\ref{fig:landau} is therefore $\varepsilon=E/\gamma-1$ for the $\Gamma$ cone and $\varepsilon=-(E/\gamma+1)$ for the H cone.
On the field side, $s=2e|B|/(\hbar c)=4\pi|\phi|/S_{\mathrm p}$ by Eqs.~(\ref{eq:flux_def}) and~(\ref{eq:comm_s}), i.e.,
\begin{equation}
s\,a^2=
\begin{cases}
64\pi\,|\phi| & (\bm{B}\parallel[100]),\\[2pt]
128\sqrt{3}\,\pi\,|\phi| & (\bm{B}\parallel[111]),
\end{cases}
\label{eq:s_phi}
\end{equation}
and the magnetic length is $\ell_B=\sqrt{\hbar c/(e|B|)}=a\sqrt{2/(s a^2)}$.
Because of the opposite charge conventions of Eqs.~(\ref{eq:H_peierls}) and~(\ref{eq:Pi_xyz}), positive $s$ corresponds to negative $\phi$ for the same spectrum; hence Figs.~\ref{fig:landau_bands} and~\ref{fig:finite_kz} are computed for $B<0$.
The range $s a^2\lesssim0.6$ of Fig.~\ref{fig:landau} thus corresponds to $\phi\lesssim3\times10^{-3}$ and $\ell_B\gtrsim1.8a$, i.e., to the immediate vicinity of $\phi=0$ in Fig.~\ref{fig:k4_Peierls}.
The linear-cone approximation requires $a^2[k_z^2+s(n+\tfrac{1}{2})]\ll1$ and thus holds for the low-lying levels only.
The highest levels of Fig.~\ref{fig:landau} ($n=8$ at $s a^2=0.6$, where $|\varepsilon|a\simeq1.1$) represent the effective model itself, and the lattice comparison of Fig.~\ref{fig:landau_bands} covers $s a^2\simeq0.08$.
At $k_z=0$ the $\sqrt{n+\tfrac{1}{2}}$ fan of Eq.~(\ref{eq:cubicLL}) is the set of Landau levels emanating from $E/\gamma=+1$ at $\phi\to0$ in Fig.~\ref{fig:k4_Peierls}, and from $E/\gamma=-1$ for the H cone, at $\phi\to0$ in Fig.~\ref{fig:k4_Peierls}(a) and at $\phi\to1/2$ in Fig.~\ref{fig:k4_Peierls}(b).
In the effective model the middle roots and the anomalous mode all lie at $E/\gamma=1$ at $k_z=0$.
On the lattice, for $\bm{B}\parallel[111]$, they form the nearly horizontal line at $E/\gamma=1$ in Fig.~\ref{fig:k4_Peierls}(e), which descends from the exactly flat sheet of Sect.~2 with the small finite-flux corrections quoted in Sect.~3.
The anomalous mode therefore becomes distinguishable only at finite $k_z$.

Figure~\ref{fig:landau_bands} tests the correspondence at finite $k_z$ by comparing the Landau levels of the effective model with the lattice spectrum of Sect.~3, computed for $\bm{B}\parallel[100]$ at the fixed flux $|\phi|=1/2400$ ($|\phi_{\mathrm{cell}}|=1/300$, $s a^2\simeq0.084$) as a function of $k_z$.
On the side of each cone that is not overlapped by the continuum of the other bands in the same $k_z$ slice [the band-4 side of the $\Gamma$ cone in Fig.~\ref{fig:landau_bands}(a) and the band-1 side of the H cone in Fig.~\ref{fig:landau_bands}(b)], the lattice levels reproduce the hyperbolic $n\geq1$ ladders, the $n=0$ pair, and the single chiral level $E=\gamma(1+k_za/2)$ [$E=-\gamma(1+k_za/2)$ for the H cone] to within about $10^{-2}\gamma$.
The residual deviations grow with $n$, as expected from the quadratic terms of the lattice dispersion neglected in Eq.~(\ref{eq:Heff}).
The anomalous level of Eq.~(\ref{eq:anomalous}) is thus a genuine feature of the lattice model, and the two cones carry chiral levels of opposite velocity, as required by Eq.~(\ref{eq:chiral}).
For the opposite sign of $B$ the lattice spectrum is mirrored, $k_z\to-k_z$.
\end{revblock}

As the flux increases beyond the linear-cone regime, the regular $\sqrt{n+\tfrac{1}{2}}$ Landau-level fan progressively breaks up and is redistributed into the miniband manifolds visible in Figs.~\ref{fig:k4_Peierls}(a) and (b).

\section{Summary}

We have investigated the electronic states of the K$_4$ lattice in a uniform magnetic field within a nearest-neighbor tight-binding model with Peierls phases.
By aligning the field with the rotated $\hat{\bm{z}}$ axis and fixing the field-parallel Bloch momentum $k_z$, the three-dimensional problem was reduced to an effective two-dimensional problem, allowing the Hofstadter spectrum to be computed for each $k_z$ as a function of the dimensionless flux \rev{$\phi = BS_{\mathrm p}/\Phi_0$ through the elementary plaquette of the projected lattice}.

For $\bm{B}\parallel[100]$, the numerical Hofstadter spectrum displays a substructure on the natural scale $\phi=1/8$ that closely resembles the square-lattice butterfly\rev{.
This scale is one flux quantum per projected unit cell, while the four-membered plaquettes tiling the (100) projection of K$_4$ set the period $\phi=1$ of the spectrum}.
For $\bm{B}\parallel[111]$, finer subpatterns emerge on the scale $\phi=1/32$ and simultaneously carry honeycomb- and triangular-lattice features, reflecting the dual character of the (111) projection\rev{.
Their alternation from window to window is controlled by the staggered Peierls phase accumulated around the elementary triangles}.
\rev{Because the elementary plaquettes are projections of the four-fold and three-fold screw helices of the chiral K$_4$ network, the field-parallel momentum enters the spectrum as an additional flux, Eq.~(\ref{eq:kz_shift}).
Together with the bipartite (chiral) symmetry of the lattice, this places the H-point triple-Dirac cone at $\phi=1/2$ in the $k_z=0$ spectrum for $\bm{B}\parallel[111]$, and it makes the spectrum symmetric about $E=0$ for every $k_z$ for $\bm{B}\parallel[100]$, so that the Landau branches of the $\Gamma$ and H cones appear as mirror images of each other.}
The connectivity of the K$_4$ network is therefore quantitatively imprinted on its magnetic energy spectrum.

In the complementary weak-field analysis, we derived the Landau levels of the triple-Dirac low-energy Hamiltonian by minimal coupling and showed that the secular equation for the three-component spinor takes the cubic form of Eq.~(\ref{eq:cubicLL}).
The cubic factorizes at $k_z=0$ into a flat branch at $\varepsilon=0$ plus a relativistic pair $\varepsilon=\pm\tfrac{1}{2}\sqrt{s(n+1/2)}$, where $\varepsilon = E/(\gamma\,a)$ denotes the rescaled eigenvalue of Eq.~(\ref{eq:E_rescaled}).
This reproduces both the pseudospin-flat zero mode characteristic of higher-pseudospin Dirac systems and the low-flux behavior of the numerical Hofstadter spectrum.
The $n=0$ sector, in which the oscillator state $|n-1\rangle$ is absent, contains an anomalous zero mode at $\varepsilon = k_z/2$ in addition to the generic pair of Eq.~(\ref{eq:n0_quadratic})\rev{.
A direct comparison with the lattice Landau bands at finite $k_z$ (Fig.~\ref{fig:landau_bands}) confirms this level as the chiral zeroth Landau level of the spin-1 node, with a counter-propagating partner at the H cone, and reproduces the hyperbolic $n\geq1$ ladders}.

These results show how the noncentrosymmetric body-centered cubic geometry, the underlying chiral octahedral group $O$, and the triple-Dirac low-energy structure of K$_4$ jointly shape its response to a magnetic field.
They provide a quantitative reference for future studies of K$_4$-type materials such as the molecular crystal of Ref.~\citen{Mizuno2015JACS} and predicted compounds based on boron, phosphorus, or hydrogenated carbon, and complement recent topological characterizations of the K$_4$ surface states.\cite{Yoshida2026PRB}
The linear-in-$k_z$ dispersion of the bulk anomalous mode $\varepsilon^{(0)}_{\mathrm{a}}=k_z/2$ contrasts with the dispersionless $n=0$ Landau level of two-dimensional Dirac fermions and is intrinsically tied to the bulk pseudospin-$1$ structure of the K$_4$ triple-Dirac cone.
\rev{How this bulk level manifests itself at the boundaries of a finite sample depends on the termination and on the field profile, and its connection to the topological charges of the triple-Dirac point and to the K$_4$ Fermi arcs of Ref.~\citen{Yoshida2026PRB} remains an open problem for future work.}

Experimentally, the most natural probe of the predicted Landau-level structure is magneto-optical absorption.\cite{Sadowski2006PRL,Jiang2007PRL,Orlita2009PRL}
In the K$_4$-type molecular crystal of Ref.~\citen{Mizuno2015JACS} and related candidate compounds, the anomalous $\varepsilon = k_z/2$ zero mode of Eq.~(\ref{eq:anomalous}) may produce a characteristic transition tied to its linear-in-$k_z$ dispersion, a feature absent in \rev{two-dimensional Dirac systems such as graphene}.

\begin{acknowledgments}
\rev{This work was supported by JSPS KAKENHI (Grants No.~JP25K01609 and No.~JP22H05473).}
\end{acknowledgments}

\appendix
\section{\rev{Hofstadter Spectra at Finite $k_z$}}
\label{app:finite_kz}

\begin{figure*}[t]
\centering
\includegraphics[width=\textwidth,clip]{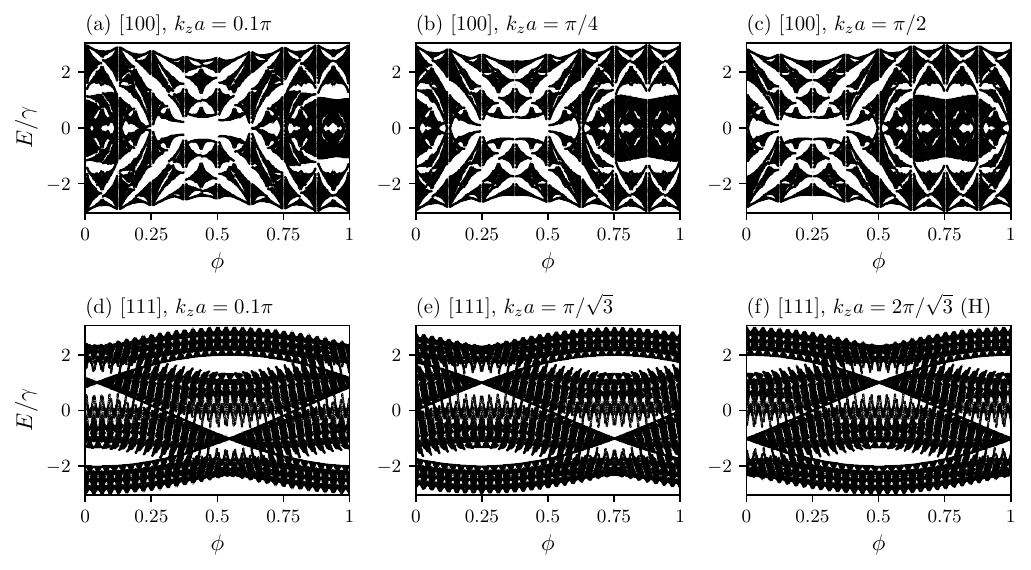}
\caption{\rev{
Hofstadter spectra of the K$_4$ lattice at finite field-parallel momentum $k_z$, computed for $B<0$ in the convention of Eqs.~(\ref{eq:H_peierls})--(\ref{eq:landau_gauge}), as in Fig.~\ref{fig:landau_bands}, and plotted as in Fig.~\ref{fig:k4_Peierls} against the magnitude of the flux $\phi$ through the elementary plaquette.
(a)--(c) $\bm{B}\parallel[100]$ at $k_za=0.1\pi$, $\pi/4$, and $\pi/2$. (d)--(f) $\bm{B}\parallel[111]$ at $k_za=0.1\pi$, $\pi/\sqrt{3}$, and $2\pi/\sqrt{3}$.
Panels (b), (c), (e), and (f) correspond to $\kappa=k_zc/2\pi=1/8$, $1/4$, $1/4$, and $1/2$, where $c$ is the lattice period along the field ($c=a$ for [100] and $c=a\sqrt{3}/2$ for [111]), for which Eq.~(\ref{eq:kz_shift}) holds exactly.
They are the $k_z=0$ spectra of Figs.~\ref{fig:k4_Peierls}(a) and (b) rigidly shifted along the flux axis, toward lower $\phi$ by $\kappa$ in (b) and (c) and toward higher $\phi$ by $\kappa$ in (e) and (f), as follows from Eq.~(\ref{eq:kz_shift}) for $B<0$.
In (f), $k_z=2\pi/(\sqrt{3}a)$ is the field-parallel momentum of the H points, and the spectrum is that of Fig.~\ref{fig:k4_Peierls}(b) inverted, $E\to-E$, in accordance with Eq.~(\ref{eq:half_shift}).
The H-point cone now appears at $E/\gamma=-1$ and $\phi=0$.
}}
\label{fig:finite_kz}
\end{figure*}

\begin{revblock}
Figure~\ref{fig:finite_kz} shows the Hofstadter spectra of Sect.~3 at finite field-parallel momentum $k_z$ for both field orientations.
The self-similar substructures and the flux scales $\phi=1/8$ and $1/32$ persist for all $k_z$, while the gaps and the triple-Dirac fans move along the flux axis.
As explained in Sect.~3, this motion is a rigid shift by $\kappa=k_zc/2\pi$ whenever $n_{\mathrm p}\kappa$ is an integer [Eq.~(\ref{eq:kz_shift}); panels (b), (c), (e), and (f)] and a continuous deformation in between [panels (a) and (d)].
For $\bm{B}\parallel[100]$ the spectrum remains symmetric about $E=0$ for every $k_z$, as required by the chiral symmetry (\ref{eq:chiral}) with $\bm{K}_{\mathrm H}\parallel\bm{B}$, so that the Landau branches derived from the $\Gamma$ and H cones appear as mirror images of each other.
For $\bm{B}\parallel[111]$, panel (f) displays the H-point cone at $\phi=0$, whose low-flux Landau fan at $E/\gamma=-1$ is the mirror image of the $\Gamma$-point fan of Fig.~\ref{fig:k4_Peierls}(e).
All spectra in this Appendix and in Fig.~\ref{fig:landau_bands} were obtained with an independent implementation of the $4q\times4q$ magnetic Bloch Hamiltonian in oblique in-plane coordinates $(u,v)$ along the projected primitive vectors, with the gauge $\bm{A}\propto u\,\nabla v$; at $\phi=0$ it reproduces Eq.~(\ref{eq:HBloch}), and its $k_z=0$ spectra coincide with Fig.~\ref{fig:k4_Peierls}.
\end{revblock}

\noindent
$^*$E-mail: WAKABAYASHI.Katsunori@nims.go.jp

\bibliographystyle{jpsj}
\bibliography{refs}

@article{Yoshida2026PRB,
  author  = {Shouya Yoshida and Katsuhiro Takahashi and Katsunori Wakabayashi},
  title   = {Topological charges, Fermi arcs, and surface states of {K4} crystal},
  journal = {Phys. Rev. B},
  volume  = {113},
  pages   = {085409},
  year    = {2026},
  doi     = {10.1103/jc96-wrmj}
}

@article{Sunada2008,
  author  = {Toshikazu Sunada},
  title   = {Crystals that nature might miss creating},
  journal = {Not. Am. Math. Soc.},
  volume  = {55},
  number  = {2},
  pages   = {208--215},
  year    = {2008}
}

@book{Sunada2013Book,
  author    = {Toshikazu Sunada},
  title     = {Topological Crystallography: With a View Towards Discrete Geometric Analysis},
  publisher = {Springer},
  address   = {Tokyo},
  year      = {2013},
  doi       = {10.1007/978-4-431-54177-6}
}

@article{Sunada2012JJM,
  author  = {Toshikazu Sunada},
  title   = {Lecture on topological crystallography},
  journal = {Jpn. J. Math.},
  volume  = {7},
  number  = {1},
  pages   = {1--39},
  year    = {2012},
  doi     = {10.1007/s11537-012-1144-4}
}

@article{OKeeffe2008,
  author  = {Michael O'Keeffe and Maxim A. Peskov and Stuart J. Ramsden and Omar M. Yaghi},
  title   = {The reticular chemistry structure resource ({RCSR}) database of, and symbols for, crystal nets},
  journal = {Acc. Chem. Res.},
  volume  = {41},
  number  = {12},
  pages   = {1782--1789},
  year    = {2008},
  doi     = {10.1021/ar800124u}
}

@article{Hyde2008,
  author  = {Stephen T. Hyde and Michael O'Keeffe and Davide M. Proserpio},
  title   = {A short history of an elusive yet ubiquitous structure in chemistry, materials, and mathematics},
  journal = {Angew. Chem. Int. Ed.},
  volume  = {47},
  number  = {42},
  pages   = {7996--8000},
  year    = {2008},
  doi     = {10.1002/anie.200801519}
}

@article{Itoh2009PRL,
  author  = {Masahiro Itoh and Motoko Kotani and Hisashi Naito and Toshikazu Sunada and Yoshiyuki Kawazoe and Tadafumi Adschiri},
  title   = {New metallic carbon crystal},
  journal = {Phys. Rev. Lett.},
  volume  = {102},
  pages   = {055703},
  year    = {2009},
  doi     = {10.1103/PhysRevLett.102.055703}
}

@article{Tsuchiizu2016PRB,
  author  = {Masahisa Tsuchiizu},
  title   = {Three-dimensional higher-spin Dirac and Weyl dispersions in the strongly isotropic {K4} crystal},
  journal = {Phys. Rev. B},
  volume  = {94},
  pages   = {195426},
  year    = {2016},
  doi     = {10.1103/PhysRevB.94.195426}
}

@article{Hagita2016JCCS,
  author  = {Katsumi Hagita and Ryoji Sahara and Yoshiyuki Kawazoe},
  title   = {{DFT} Study of Carbon Crystals from Nested {K4} Lattice to {BC}-8 Structure},
  journal = {J. Chin. Chem. Soc.},
  volume  = {63},
  number  = {6},
  pages   = {526--532},
  year    = {2016},
  doi     = {10.1002/jccs.201500376}
}

@article{Dai2010PCCP,
  author  = {Jun Dai and Zhenyu Li and Jinlong Yang},
  title   = {Boron {K4} crystal: a stable chiral three-dimensional sp$^2$ network},
  journal = {Phys. Chem. Chem. Phys.},
  volume  = {12},
  number  = {39},
  pages   = {12420--12422},
  year    = {2010},
  doi     = {10.1039/C0CP00735H}
}

@article{Liu2016SR,
  author  = {Jie Liu and Shunhong Zhang and Yaguang Guo and Qian Wang},
  title   = {Phosphorus {K4} Crystal: A New Stable Allotrope},
  journal = {Sci. Rep.},
  volume  = {6},
  pages   = {37528},
  year    = {2016},
  doi     = {10.1038/srep37528}
}

@article{Lian2013JCP,
  author  = {C.-S. Lian and X.-Q. Wang and J.-T. Wang},
  title   = {Hydrogenated {K4} carbon: A new stable cubic gauche structure of carbon hydride},
  journal = {J. Chem. Phys.},
  volume  = {138},
  number  = {2},
  pages   = {024702},
  year    = {2013},
  doi     = {10.1063/1.4773584}
}

@article{Wen2011CPL,
  author  = {Bin Wen and Seiichi Takami and Yoshiyuki Kawazoe and Tadafumi Adschiri},
  title   = {Carbon-doped {K4} nitrogen: A novel high energy density material},
  journal = {Chem. Phys. Lett.},
  volume  = {506},
  number  = {4--6},
  pages   = {175--178},
  year    = {2011},
  doi     = {10.1016/j.cplett.2011.02.042}
}

@article{Wen2012JPCS,
  author  = {Bin Wen and Seiichi Takami and Yoshiyuki Kawazoe and Tadafumi Adschiri},
  title   = {Mechanical stabilities of {K4} carbon and {K4}-like {NaC2}},
  journal = {J. Phys. Chem. Solids},
  volume  = {73},
  number  = {11},
  pages   = {1264--1267},
  year    = {2012},
  doi     = {10.1016/j.jpcs.2012.07.013}
}

@article{Mizuno2015JACS,
  author  = {Asato Mizuno and Yoshiaki Shuku and Rie Suizu and Michio M. Matsushita and Masahisa Tsuchiizu and Daniel Reta Ma{\~n}eru and Francesc Illas and Vincent Robert and Kunio Awaga},
  title   = {Discovery of the {K4} Structure Formed by a Triangular $\pi$ Radical Anion},
  journal = {J. Am. Chem. Soc.},
  volume  = {137},
  number  = {24},
  pages   = {7612--7615},
  year    = {2015},
  doi     = {10.1021/jacs.5b04180}
}

@article{Bradlyn2016Science,
  author  = {Barry Bradlyn and Jennifer Cano and Zhijun Wang and M. G. Vergniory and Claudia Felser and R. J. Cava and B. Andrei Bernevig},
  title   = {Beyond Dirac and Weyl fermions: Unconventional quasiparticles in conventional crystals},
  journal = {Science},
  volume  = {353},
  number  = {6299},
  pages   = {aaf5037},
  year    = {2016},
  doi     = {10.1126/science.aaf5037}
}

@article{Armitage2018RMP,
  author  = {N. P. Armitage and E. J. Mele and Ashvin Vishwanath},
  title   = {Weyl and Dirac semimetals in three-dimensional solids},
  journal = {Rev. Mod. Phys.},
  volume  = {90},
  pages   = {015001},
  year    = {2018},
  doi     = {10.1103/RevModPhys.90.015001}
}

@article{Peierls1933ZPhys,
  author  = {Rudolf Peierls},
  title   = {Zur Theorie des Diamagnetismus von Leitungselektronen},
  journal = {Z. Phys.},
  volume  = {80},
  number  = {11--12},
  pages   = {763--791},
  year    = {1933},
  doi     = {10.1007/BF01342591}
}

@article{Landau1930ZPhys,
  author  = {L. Landau},
  title   = {Diamagnetismus der Metalle},
  journal = {Z. Phys.},
  volume  = {64},
  number  = {9--10},
  pages   = {629--637},
  year    = {1930},
  doi     = {10.1007/BF01397213}
}

@article{Harper1955,
  author  = {P. G. Harper},
  title   = {Single Band Motion of Conduction Electrons in a Uniform Magnetic Field},
  journal = {Proc. Phys. Soc. A},
  volume  = {68},
  number  = {10},
  pages   = {874--878},
  year    = {1955},
  doi     = {10.1088/0370-1298/68/10/304}
}

@article{Brown1964PR,
  author  = {E. Brown},
  title   = {Bloch Electrons in a Uniform Magnetic Field},
  journal = {Phys. Rev.},
  volume  = {133},
  number  = {4A},
  pages   = {A1038--A1044},
  year    = {1964},
  doi     = {10.1103/PhysRev.133.A1038}
}

@article{Zak1964PR,
  author  = {J. Zak},
  title   = {Magnetic Translation Group},
  journal = {Phys. Rev.},
  volume  = {134},
  number  = {6A},
  pages   = {A1602--A1606},
  year    = {1964},
  doi     = {10.1103/PhysRev.134.A1602}
}

@article{Hofstadter1976PRB,
  author  = {Douglas R. Hofstadter},
  title   = {Energy levels and wave functions of {Bloch} electrons in rational and irrational magnetic fields},
  journal = {Phys. Rev. B},
  volume  = {14},
  number  = {6},
  pages   = {2239--2249},
  year    = {1976},
  doi     = {10.1103/PhysRevB.14.2239}
}

@article{Thouless1982PRL,
  author  = {D. J. Thouless and M. Kohmoto and M. P. Nightingale and M. den Nijs},
  title   = {Quantized Hall Conductance in a Two-Dimensional Periodic Potential},
  journal = {Phys. Rev. Lett.},
  volume  = {49},
  number  = {6},
  pages   = {405--408},
  year    = {1982},
  doi     = {10.1103/PhysRevLett.49.405}
}

@article{Kohmoto1985AnnPhys,
  author  = {Mahito Kohmoto},
  title   = {Topological invariant and the quantization of the Hall conductance},
  journal = {Ann. Phys.},
  volume  = {160},
  number  = {2},
  pages   = {343--354},
  year    = {1985},
  doi     = {10.1016/0003-4916(85)90148-4}
}

@article{Hasegawa1989PRL,
  author  = {Y. Hasegawa and P. Lederer and T. M. Rice and P. B. Wiegmann},
  title   = {Theory of Electronic Diamagnetism in Two-Dimensional Lattices},
  journal = {Phys. Rev. Lett.},
  volume  = {63},
  number  = {8},
  pages   = {907--910},
  year    = {1989},
  doi     = {10.1103/PhysRevLett.63.907}
}

@article{Janecek2013PRB,
  author  = {S. Jane{\v c}ek and M. Aichinger and E. R. Hern{\'a}ndez},
  title   = {Two-dimensional {Bloch} electrons in perpendicular magnetic fields: An exact calculation of the Hofstadter butterfly spectrum},
  journal = {Phys. Rev. B},
  volume  = {87},
  pages   = {235429},
  year    = {2013},
  doi     = {10.1103/PhysRevB.87.235429}
}

@article{Wallace1947PR,
  author  = {P. R. Wallace},
  title   = {The Band Theory of Graphite},
  journal = {Phys. Rev.},
  volume  = {71},
  number  = {9},
  pages   = {622--634},
  year    = {1947},
  doi     = {10.1103/PhysRev.71.622}
}

@article{Slonczewski1958PR,
  author  = {J. C. Slonczewski and P. R. Weiss},
  title   = {Band Structure of Graphite},
  journal = {Phys. Rev.},
  volume  = {109},
  number  = {2},
  pages   = {272--279},
  year    = {1958},
  doi     = {10.1103/PhysRev.109.272}
}

@article{McClure1956PR,
  author  = {J. W. McClure},
  title   = {Diamagnetism of Graphite},
  journal = {Phys. Rev.},
  volume  = {104},
  number  = {3},
  pages   = {666--671},
  year    = {1956},
  doi     = {10.1103/PhysRev.104.666}
}

@article{McClure1957PR,
  author  = {J. W. McClure},
  title   = {Band Structure of Graphite and de Haas--van Alphen Effect},
  journal = {Phys. Rev.},
  volume  = {108},
  number  = {3},
  pages   = {612--618},
  year    = {1957},
  doi     = {10.1103/PhysRev.108.612}
}

@article{McClure1960PR,
  author  = {J. W. McClure},
  title   = {Theory of Diamagnetism of Graphite},
  journal = {Phys. Rev.},
  volume  = {119},
  number  = {2},
  pages   = {606--613},
  year    = {1960},
  doi     = {10.1103/PhysRev.119.606}
}

@article{Dresselhaus1974PRB,
  author  = {G. Dresselhaus},
  title   = {Graphite Landau levels in the presence of trigonal warping},
  journal = {Phys. Rev. B},
  volume  = {10},
  number  = {8},
  pages   = {3602--3609},
  year    = {1974},
  doi     = {10.1103/PhysRevB.10.3602}
}

@article{Nakao1976JPSJ,
  author  = {Kenji Nakao},
  title   = {Landau Level Structure and Magnetic Breakthrough in Graphite},
  journal = {J. Phys. Soc. Jpn.},
  volume  = {40},
  number  = {3},
  pages   = {761--768},
  year    = {1976},
  doi     = {10.1143/JPSJ.40.761}
}

@article{McCann2006PRL,
  author  = {Edward McCann and Vladimir I. Fal'ko},
  title   = {Landau-Level Degeneracy and Quantum Hall Effect in a Graphite Bilayer},
  journal = {Phys. Rev. Lett.},
  volume  = {96},
  pages   = {086805},
  year    = {2006},
  doi     = {10.1103/PhysRevLett.96.086805}
}

@article{Sadowski2006PRL,
  author  = {M. L. Sadowski and G. Martinez and M. Potemski and C. Berger and W. A. de Heer},
  title   = {Landau Level Spectroscopy of Ultrathin Graphite Layers},
  journal = {Phys. Rev. Lett.},
  volume  = {97},
  pages   = {266405},
  year    = {2006},
  doi     = {10.1103/PhysRevLett.97.266405}
}

@article{Jiang2007PRL,
  author  = {Z. Jiang and E. A. Henriksen and L.-C. Tung and Y.-J. Wang and M. E. Schwartz and M. Y. Han and P. Kim and H. L. Stormer},
  title   = {Infrared Spectroscopy of Landau Levels of Graphene},
  journal = {Phys. Rev. Lett.},
  volume  = {98},
  pages   = {197403},
  year    = {2007},
  doi     = {10.1103/PhysRevLett.98.197403}
}

@article{CastroNeto2009RMP,
  author  = {A. H. Castro Neto and F. Guinea and N. M. R. Peres and K. S. Novoselov and A. K. Geim},
  title   = {The electronic properties of graphene},
  journal = {Rev. Mod. Phys.},
  volume  = {81},
  pages   = {109--162},
  year    = {2009},
  doi     = {10.1103/RevModPhys.81.109}
}

@article{Orlita2009PRL,
  author  = {M. Orlita and C. Faugeras and J. M. Schneider and G. Martinez and D. K. Maude and M. Potemski},
  title   = {Graphite from the Viewpoint of Landau Level Spectroscopy: An Effective Graphene Bilayer and Monolayer},
  journal = {Phys. Rev. Lett.},
  volume  = {102},
  pages   = {166401},
  year    = {2009},
  doi     = {10.1103/PhysRevLett.102.166401}
}

@article{Goerbig2011RMP,
  author  = {Mark O. Goerbig},
  title   = {Electronic properties of graphene in a strong magnetic field},
  journal = {Rev. Mod. Phys.},
  volume  = {83},
  pages   = {1193--1243},
  year    = {2011},
  doi     = {10.1103/RevModPhys.83.1193}
}

@article{Dean2013Nature,
  author  = {C. R. Dean and L. Wang and P. Maher and C. Forsythe and F. Ghahari and Y. Gao and J. Katoch and M. Ishigami and P. Moon and M. Koshino and T. Taniguchi and K. Watanabe and K. L. Shepard and J. Hone and P. Kim},
  title   = {Hofstadter's butterfly and the fractal quantum {Hall} effect in moir{\'e} superlattices},
  journal = {Nature},
  volume  = {497},
  pages   = {598--602},
  year    = {2013},
  doi     = {10.1038/nature12186}
}

@article{Ponomarenko2013Nature,
  author  = {L. A. Ponomarenko and R. V. Gorbachev and G. L. Yu and D. C. Elias and R. Jalil and A. A. Patel and A. Mishchenko and A. S. Mayorov and C. R. Woods and J. R. Wallbank and M. Mucha-Kruczy{\'n}ski and B. A. Piot and M. Potemski and I. V. Grigorieva and K. S. Novoselov and F. Guinea and V. I. Fal'ko and A. K. Geim},
  title   = {Cloning of Dirac fermions in graphene superlattices},
  journal = {Nature},
  volume  = {497},
  pages   = {594--597},
  year    = {2013},
  doi     = {10.1038/nature12187}
}

@article{McCann2013RPP,
  author  = {Edward McCann and Mikito Koshino},
  title   = {The electronic properties of bilayer graphene},
  journal = {Rep. Prog. Phys.},
  volume  = {76},
  number  = {5},
  pages   = {056503},
  year    = {2013},
  doi     = {10.1088/0034-4885/76/5/056503}
}

@article{Raoux2014PRL,
  author  = {A. Raoux and M. Morigi and J.-N. Fuchs and F. Pi\'echon and G. Montambaux},
  title   = {From Dia- to Paramagnetic Orbital Susceptibility of Massless Fermions},
  journal = {Phys. Rev. Lett.},
  volume  = {112},
  pages   = {026402},
  year    = {2014},
  doi     = {10.1103/PhysRevLett.112.026402}
}

@article{Malcolm2014PRB,
  author  = {J. D. Malcolm and E. J. Nicol},
  title   = {Magneto-optics of general pseudospin-$s$ two-dimensional Dirac-Weyl fermions},
  journal = {Phys. Rev. B},
  volume  = {90},
  pages   = {035405},
  year    = {2014},
  doi     = {10.1103/PhysRevB.90.035405}
}

@article{Illes2016PRB,
  author  = {E. Illes and E. J. Nicol},
  title   = {Magnetic Properties of the $\alpha$-${T}_{3}$ Model: Magneto-Optical Conductivity and the Hofstadter Butterfly},
  journal = {Phys. Rev. B},
  volume  = {94},
  pages   = {125435},
  year    = {2016},
  doi     = {10.1103/PhysRevB.94.125435}
}

@article{KoshinoAoki2001PRL,
  author  = {Mikito Koshino and Hideo Aoki and Kazuhiko Kuroki and Seiichi Kagoshima and Toshihito Osada},
  title   = {Hofstadter Butterfly and Integer Quantum {Hall} Effect in Three Dimensions},
  journal = {Phys. Rev. Lett.},
  volume  = {86},
  number  = {6},
  pages   = {1062--1065},
  year    = {2001},
  doi     = {10.1103/PhysRevLett.86.1062}
}

@article{Weng2016PRB,
  author  = {Hongming Weng and Chen Fang and Zhong Fang and Xi Dai},
  title   = {Coexistence of {Weyl} fermion and massless triply degenerate nodal points},
  journal = {Phys. Rev. B},
  volume  = {94},
  pages   = {165201},
  year    = {2016},
  doi     = {10.1103/PhysRevB.94.165201}
}

@article{Lv2017Nature,
  author  = {B. Q. Lv and Z.-L. Feng and Q.-N. Xu and X. Gao and J.-Z. Ma and L.-Y. Kong and P. Richard and Y.-B. Huang and V. N. Strocov and C. Fang and H.-M. Weng and Y.-G. Shi and T. Qian and H. Ding},
  title   = {Observation of three-component fermions in the topological semimetal molybdenum phosphide},
  journal = {Nature},
  volume  = {546},
  pages   = {627--631},
  year    = {2017},
  doi     = {10.1038/nature22390}
}

@article{Burkov2011PRL,
  author  = {A. A. Burkov and Leon Balents},
  title   = {{Weyl} Semimetal in a Topological Insulator Multilayer},
  journal = {Phys. Rev. Lett.},
  volume  = {107},
  pages   = {127205},
  year    = {2011},
  doi     = {10.1103/PhysRevLett.107.127205}
}

@article{Lv2015PRX,
  author  = {B. Q. Lv and H. M. Weng and B. B. Fu and X. P. Wang and H. Miao and J. Ma and P. Richard and X. C. Huang and L. X. Zhao and G. F. Chen and Z. Fang and X. Dai and T. Qian and H. Ding},
  title   = {Experimental Discovery of {Weyl} Semimetal {TaAs}},
  journal = {Phys. Rev. X},
  volume  = {5},
  pages   = {031013},
  year    = {2015},
  doi     = {10.1103/PhysRevX.5.031013}
}

@article{Sutherland1986PRB,
  author  = {Bill Sutherland},
  title   = {Localization of electronic wave functions due to local topology},
  journal = {Phys. Rev. B},
  volume  = {34},
  number  = {8},
  pages   = {5208--5211},
  year    = {1986},
  doi     = {10.1103/PhysRevB.34.5208}
}
\end{document}